\documentclass[letterpaper]{aa} 
\usepackage{graphicx}
\usepackage{natbib}
\bibpunct{(}{)}{;}{a}{}{,}
\usepackage{txfonts}
\usepackage{aalongtable}

\begin{document}
\title{Hipparcos preliminary astrometric masses for the two close-in companions to \object{HD 131664} and \object{HD 43848}
}
   \subtitle{A brown dwarf and a low mass star}

\author{A. Sozzetti\inst{1} \and S. Desidera\inst{2}}

\offprints{A. Sozzetti, \\\email{sozzetti@oato.inaf.it}}

\institute{INAF - Osservatorio Astronomico di Torino, via Osservatorio 20,
     I-10025 Pino Torinese, Italy
\and INAF - Osservatorio Astronomico di Padova, Vicolo dell'Osservatorio 5, I-35122 Padova, Italy}

\date{Received ......; accepted ......}

  \abstract
    {Several mechanisms for forming brown dwarfs have
    been proposed, which are today believed not to be mutually exclusive.
    Among the fundamental characteristics of brown dwarfs which are intrinsically tied to their
    origin, multiplicity is of particular relevance.
    Any successful determination of the actual mass for such objects in systems
    is thus worthwhile as it allows one to improve on the characterization of the multiplicity
    properties (e.g, frequency, separation, mass-ratio distribution) of sub-stellar companions. }
   {We attempt to improve on the characterization of the properties (orbital
   elements, masses) of two Doppler-detected sub-stellar companions to the nearby
   G dwarfs \object{HD 131664} and \object{HD 43848}.}
   {We carry out orbital fits to the Hipparcos Intermediate Astrometric Data (IAD) for the two
   stars, taking advantage of the knowledge of the spectroscopic orbits, and solving for the
   two orbital elements that can be determined in principle solely by astrometry, the
   inclination angle $i$ and the longitude of the ascending node $\Omega$. A number of
   checks are carried out in order to assess the reliability of the orbital solutions thus obtained.}
   {The best-fit solution for \object{HD 131664} yields $i=55\pm33$ deg and $\Omega=22\pm28$ deg.
   The resulting inferred true companion mass is then $M_c = 23_{-5}^{+26}$ $M_J$. For \object{HD 43848}, we find
   $i=12\pm7$ deg and $\Omega=288\pm22$ deg, and a corresponding $M_c = 120_{-43}^{+167}$ $M_J$.
   Based on the statistical evidence from an $F$-test, the study of the joint confidence intervals of variation
   of $i$ and $\Omega$, and the comparison of the derived orbital semi-major axes with a distribution
   of false astrometric orbits obtained for single stars observed by Hipparcos, the astrometric signal of
   the two companions to \object{HD 131664} and \object{HD 43848} is then considered detected in the Hipparcos IAD, with a
   level of statistical confidence not exceeding 95\%.  }
   {We constrain the true mass of \object{HD 131664}b to that of a brown dwarf to within a somewhat
   statistically significant degree of confidence ($\sim2-\sigma$). For \object{HD 43848}b, a true mass in the
   brown dwarf regime is ruled out at the $1-\sigma$ confidence level.
   The results are discussed in the context of the properties of the (few) close sub-stellar and
   massive planetary companions to nearby solar-type stars and their implications for proposed models of
   formation and structure of massive planets and brown dwarfs.}


\titlerunning{Hipparcos preliminary astrometric masses for the companions
to \object{HD 131664} and \object{HD 43848}}
\authorrunning{A. Sozzetti \& S. Desidera}
\maketitle

%

\section{Introduction}

Similarly to the detection of the first Jupiter-mass ($M_J$) planet orbiting a star other
than the Sun~\citep{mayor95}, the first unambiguous discovery of a brown
dwarf\footnote{We operationally adopt the commonly used definition of brown dwarf
as a deuterium-burning sub-stellar object with mass in the approximate range $15-80$
$M_J$} dates back $\sim14$ years~\citep[e.g.,][]{nakajima95,rebolo95}.
Several mechanisms for forming brown dwarfs have
been proposed: Turbulent fragmentation of molecular clouds, fragmentation of
massive prestellar cores, protoplanetary disk fragmentation, dynamical ejection of
protostellar embryos, and photoerosion of protostellar cores~\citep[for a review, see][]{whit07}.
Such mechanisms are today believed not to be mutually exclusive, and they all
likely operate in nature. To determine their relative contribution to the
overall brown dwarf population is a major theoretical challenge. To date,
agreement is still to be reached on fundamental issues such as the minimum mass for brown dwarfs, and
how brown dwarfs might be distinguished from planets. For example, if brown dwarfs
are identified as objects that form like stars do, on dynamical timescales by
gravitational instability, regardless of the formation locus (a molecular cloud or a
marginally unstable protoplanetary disk) and if the minimum mass for core collapse is a few $M_J$
~\citep[see][]{whit07}, then significant overlap between the mass range of brown dwarfs
and planets would occur, and a simple mass cutoff (such as the one adopted here)
may not apply. Given the significant number of open questions in this research field,
it is thus essential for theory to accurately reproduce, and for observations to
carefully determine, the ensemble properties of brown dwarfs,
including the brown dwarf initial mass function, the young brown dwarfs
kinematics, distribution, and disk frequency, and the binary statistics of brown
dwarfs across a wide range of primary masses and orbital separations (for a review
see~\citealt{luhman07} and~\citealt{burga07}).

Among the fundamental characteristics of brown dwarfs which are
intrinsically tied to their origin, multiplicity properties (e.g,
frequency, separation, mass-ratio distribution) are of particular
relevance, as together with the possibility of determining the
actual internal composition and atmospheric features of individual
objects, they constitute one of the few ways to observationally
distinguish between planets and brown dwarfs in the possible
overlap region in mass. For example, the frequency of close ($a<5$
AU) stellar ($M_c>0.08$ $M_\odot$) companions to nearby ($d<50$
pc) solar-type stars is $13\pm3$\%~\citep{duque91}, while in the
same range of separations the frequency of giant planets ($M_c
\lesssim 15$ $M_J$) is known today to be about
7\%~\citep{marcy08}. Brown dwarfs, on the other hand, appear
conspicuously absent in the datasets collected by decade-long,
high-precision radial-velocity surveys of thousands of bright
normal stars, despite the fact that their large RV signals would
have been easily spotted. Initial claims that the frequency of
close brown dwarf companions seemed to be in fair agreement with a
constant distribution of mass ratios~\citep{mayor92} where later
dismissed by studies that showed, based on a combination of radial
velocity measurements and Hipparcos astrometric observations, how
most of these putative sub-stellar companions where in fact
stars~\citep{halb00}. Indeed, among close companions, brown dwarfs
appear outnumbered by stars and planets by factors of $\sim100$
and $\sim50$, respectively, with typical frequency estimates of
$\approx0.1$\%~\citep{marcy00}. Only about half a dozen close
companions with minimum masses in the brown dwarf regime are known
today around bright, nearby solar-type stars It is conceivable
that observational biases might somewhat contribute to a reduction
of the discovery rate of brown dwarfs with respect to planets
(massive companions being assigned typically lower priorities than
lower-mass planets in Doppler surveys hard-pressed for optimal use
of the precious observing time at 10-m class telescopes). However,
such biases do not seem capable of explaining two orders of
magnitude of difference in the observed frequency of brown dwarf
companions with respect to planets and stellar companions.

The dearth of close brown dwarf companions to solar-type stars, commonly referred to as
the 'brown dwarf desert'~\citep[e.g,][]{campbell88,marcy00},
is seen to extend at wider separations. Only recently~\citet{patel07} have reported
the first radial velocity detection of a handful of companions with $M_c\sin i$ firmly
established in the brown dwarf mass range at orbital distances of $\sim4$ to 18 AU.
These first results do not allow yet to produce an actual number for the brown dwarf frequency
in this separation range. At larger separations ($\sim 50-1000$ AU), near-IR direct imaging surveys
have confirmed a deficit in sub-stellar companions relative to stellar companions, but not quite
as extreme as that apparent at orbital radii within a few AUs.
~\citet{mccarthy04} find $f_\mathrm{BD} = 1\pm1$\%, roughly a factor of ten lower
than the stellar companion frequency in the same separation range. More recently
~\citet{metchev09} have derived $f_\mathrm{BD} = 3.2^{+3.1}_{-2.7}$\%, a number
formally compatible with that of~\citet{mccarthy04}. In any case, the abundance
of wide-separation brown dwarf companions is comparable to that of free-floating brown dwarfs relative to
stars. The evidence for a not-so-dry desert at wide separations is usually interpreted as supporting
the view that brown dwarfs form by core fragmentation just like stars. However, the finer details of the
formation mechanism are not well understood, as mentioned above, and a more accurate characterization of
the multiplicity of brown dwarfs is needed in order to shed light on the relative role of the various
proposed scenarios.

The dynamical determination of the mass of a few close brown dwarfs companions to low-mass stars and
of brown dwarf binaries has been obtained for eclipsing systems for which both spectra can be observed
~\citep[e.g.,][]{zapatero04,stassun06}, and by a combination of aperture masking interferometry
and astrometry~\citep{ireland08}. As for the sample of companions to nearby dwarfs discovered
by Doppler planet surveys with $M_c\sin i$ values in the range between high-mass planets and brown dwarfs,
it is conceivable that some of them are in reality stars seen pole-on. Similarly to~\citet{halb00},
several authors have attempted to combine the radial-velocity orbits with Hipparcos astrometry in order
to determine the inclination and true mass of the companions (for a review see for example~\citealt{sozzetti09}).
Recently, ~\citet{reffert06}
presented low-significance detections of the astrometric orbits of the two outer companions in the
HD 38529 and HD 168443 planetary systems, inferring masses in the brown dwarf regime
of 37 $M_J$ and 34 $M_J$ for HD 38529c and HD 168443c, respectively. High-precision astrometry with
HST/FGS allowed~\citet{bean07} to determine an actual mass of 0.14 $M_\odot$ for the companion to
HD 33636, originally published with $M_c\sin i = 9.3$ $M_J$. Any successful attempt to resolve the ambiguity in
the actual companion mass for such objects is thus worthwhile as it allows one to improve on the
characterization of the multiplicity of brown dwarfs. This in turn permits to better understand the global
nature of the sub-stellar companions found at larger separations by direct imaging surveys.

In this paper we present a new attempt at combining the
information from Doppler measurements with Hipparcos astrometry to
better the constraints on the mass of two Doppler-detected
low-mass companions to HD 131664~\citep{moutou09} and \object{HD
43848}~\citep{minniti09}, with published minimum masses in the
low-mass brown dwarf regime. A summary of the available data in
the literature for the two systems is presented in \S~\ref{summ}.
We derive improved contraints on the actual companion masses in
\S~\ref{astrom}. Finally, in \S~\ref{discuss} we a) put the new
results in the context of the properties of the (few) close
sub-stellar and massive planetary companions to nearby solar-type
stars, collected in a catalogue table for ease of consultation and
reference in future works, and b) discuss some of the implications
of the present-day observational evidence for formation and
structural models of massive planets and brown dwarfs.

\section{Doppler data and Hipparcos astrometry}\label{summ}

\begin{table*}
\caption{Stellar characteristics of \object{HD 131664} and \object{HD 43848}
and spectroscopic orbital elements for the two Doppler-detected companions.
Reported errors are from the discovery papers.}
\label{parsum}      
\centering                          
\begin{tabular}{c c c c c c c c c c c c c}        
\hline\hline                 
Star Name & Sp.T. & $V$ & $\pi_\star$ & $d$& $M_\star$  & $P$ & $T_0$ & $e$ & $\omega$ & $K$ & $M_c\sin i$ & $a_1$\\    
&  & (mag) & (mas) & (pc) & ($M_\odot$) & (d) & (JD-2400000) & & (deg) & (m s$^{-1}$) & $(M_J)$ & (AU) \\
\hline
 & & & & & & & & & & & & \\
HD 131664 & G3V & 8.13 & 18.04 & 55.43 & 1.10       & 1951     & 52060   & 0.638      & 149.7    & 359.5     & 18.15     & 3.17 \\      
          &     &      &       &       & $\pm0.03$  & $\pm41$  & $\pm41$ & $\pm0.02$  & $\pm1.0$ & $\pm22.3$ & $\pm0.35$ &      \\
 & & & & & & & & & & & & \\
\hline
 & & & & & & & & & & & & \\
HD 43848 & G2V & 8.65 & 26.99 & 37.05 & 0.89 & 2371     & 53227   & 0.690     & 229.0    & 544      &   25.0 &  3.40 \\      
         &     &      &       &       &      & $\pm840$ & $\pm65$ & $\pm0.12$ & $\pm9.0$ & $\pm200$ &        &       \\
 & & & & & & & & & & & & \\
\hline
\end{tabular}
\end{table*}

\subsection{HD 131664}

Included in the Doppler search for southern extrasolar planets carried out with the HARPS
~\citep{pepe03} spectrograph on the ESO 3.6-m telescope at La Silla Observatory,
the bright G-type star \object{HD 131664} (HIP 73408) was recently announced by~\citet{moutou09} to
be orbited by a companion with minimum mass of $M_c\sin i = 18$ $M_J$ on an eccentric
orbit of period $P\approx5.3$ yr (for convenience, see Table~\ref{parsum} for a summary of the properties
of the primary and of the orbital parameters of detected companion). At the distance of \object{HD 131664},
the inferred orbital separation ranges between 35 mas and 100 mas (as pointed out by Moutou et al.),
making it a potentially interesting target for future direct imaging observations. The viability of
this investigation depends in particular on the true mass of the companion and the actual contrast ratio.
In the former case, high-precision astrometry with Gaia in space and VLTI/PRIMA from the ground
~\citep[e.g., ][and references therein]{sozzetti09}
will allow to derive accurate values of the actual mass of the companion. However, it must be noted that
the minimum astrometric signature induced on the primary is just under 1 mas, as opposed to a median
single-measurement error $\sigma_\mathrm{HIP}\approx3$ mas. Hipparcos observations of this
star could then help place useful mass constraints on the companion mass. Indeed, the Double and Multiple
Star Annex of the Hipparcos Catalogue reports a G flag, indicating that a 7-parameter solution
(allowing for acceleration in the proper motion) was found to significantly improve the standard
5-parameter single-star fit. The reported acceleration solution for \object{HD 131664} is based on 96 datapoints,
two of which corresponding to observations retained only by the NDAC consortium.

\subsection{HD 43848}

Doppler measurements taken with the MIKE echelle
spectrograph~\citep{bernstein03} on the 6.5-m Magellan II (Clay)
telescope have revealed~\citep{minniti09} the nearby solar-type
star \object{HD 43848} (HIP 29804) to be orbited by a companion
with $M_c\sin i = 25$ $M_J$ on a high-eccentricity orbit with
$P\approx6.5$ yr (for completeness, Table~\ref{parsum} also
reports the summary of the properties of the primary and of the
orbital parameters of detected companion). This system also
appears to be of potential interest for future direct imaging
observations, with a separation at apoastron of $\approx 0.2$
arcsec. Its minimum astrometric signature, at the distance of HD
43848, is $\approx 2.5$ mas, a value comparable to the typical
precision of Hipparcos astrometry for this star (median error
$\sigma_\mathrm{HIP}\sim3.3$ mas). An astrometric solution with
acceleration terms for this star is also present in the Hipparcos
Catalogue, based on a total of 76 measurements \footnote{Note
that~\citet{minniti09} looked for, and failed to find, hints of
higher dispersion in the Hipparcos measurements for \object{HD
43848} with respect to those of stars of similar magnitude and
distance.} (with one abscissa retained only by FAST, one only by
NDAC, and one rejected by NDAC in the solution). For both
\object{HD 43848} and \object{HD 131664} there appears to be
indication of the presence of a long-period trend in the Hipparcos
data, rendering worthwhile a further investigation of the
available astrometry.

\section{Combined radial velocity+astrometry solution}\label{astrom}

When searching for evidence of the presence of orbital signal in Hipparcos data of a
given star due to a spectroscopically discovered low-mass companion, a typical procedure is
applied in which information
from radial velocities is assumed known, and one resorts to probe the region of the parameter
space not covered by spectroscopy. Orbital fits to the Hipparcos IAD are then
usually performed~\citep{mazeh99,halb00,han01,zucker00,reffert06,witten09}
by keeping fixed four orbital elements ($P$, $e$, $T_0$, $\omega$) to their spectroscopically
determined values and by solving for inclination angle $i$ and position angle
of the ascending node $\Omega$, with the additional constraint that the astrometric
semi-major axis satisfies the equality~\citep{pourbaix00}:

\begin{equation}\label{constraint}
a\sin i = 9.19\times 10^{-8}P K\sqrt{1 - e^2}\pi_\star\,\,\,\mathrm{mas},
\end{equation}

\noindent
where $P$ is in days, the semi-amplitude of the radial velocity curve $K$ is in m s$^{-1}$,
and the orbital parallax $\pi_\star$ is in mas. The resulting fitting procedure has then a total of
7 adjustable parameters, i.e. the five astrometric parameters (two position offsets in
right ascension and declination, two differential corrections to the proper motion in
right ascension and declination, and the parallax) plus $i$ and $\Omega$.

The experience of various authors indicates that one must be careful in not asking Hipparcos
data to tell more than they actually can.~\citet{pourbaix01},~\citet{pourbare01}, and
later~\citet{zuck01} have for example shown that the Hipparcos IAD, while useful to put upper limits on the
size of the astrometric perturbations, must be interpreted with great caution when attempting
to derive actual astrometric orbits for sub-stellar companions with semi-major axes close to
or even below the typical single-measurement precision of the satellite. So while the astrometric
orbit does not need to be actually detected in order to derive constraints on the values of
$i$ and $M_c$, when not outright refuted~\citep[e.g., ][]{pourbare01},
reported detections usually have relatively low levels of statistical confidence~\citep[e.g., ][]{reffert06}.
Based on the above considerations, we have adopted a multi-step
approach to the treatment of the Hipparcos IAD for \object{HD 131664} and \object{HD 43848}
in an attempt to make a statistically solid case for any conclusions we derive.

   \begin{figure}
   \centering
$\begin{array}{c}
\includegraphics[width=0.45\textwidth]{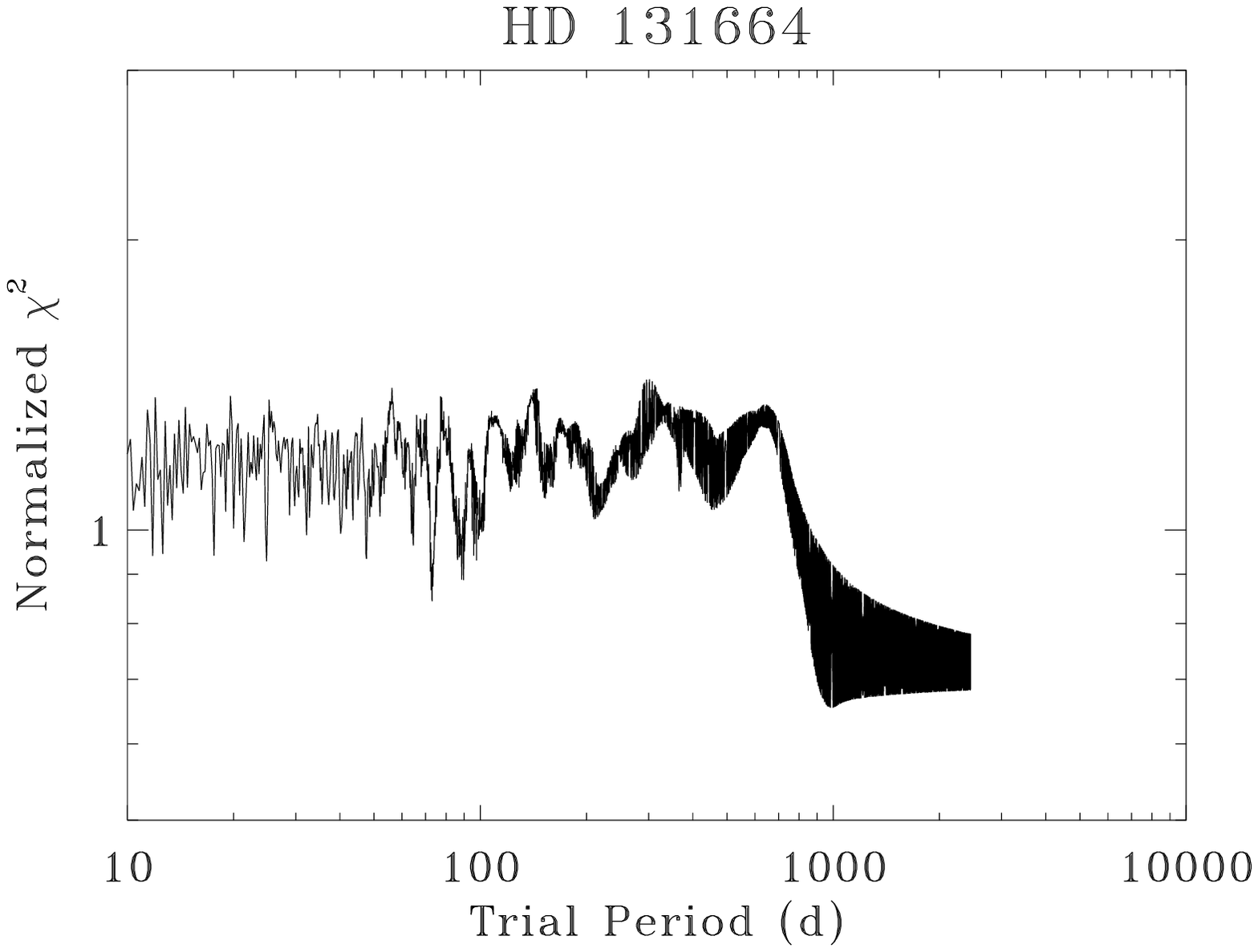} \\
\includegraphics[width=0.45\textwidth]{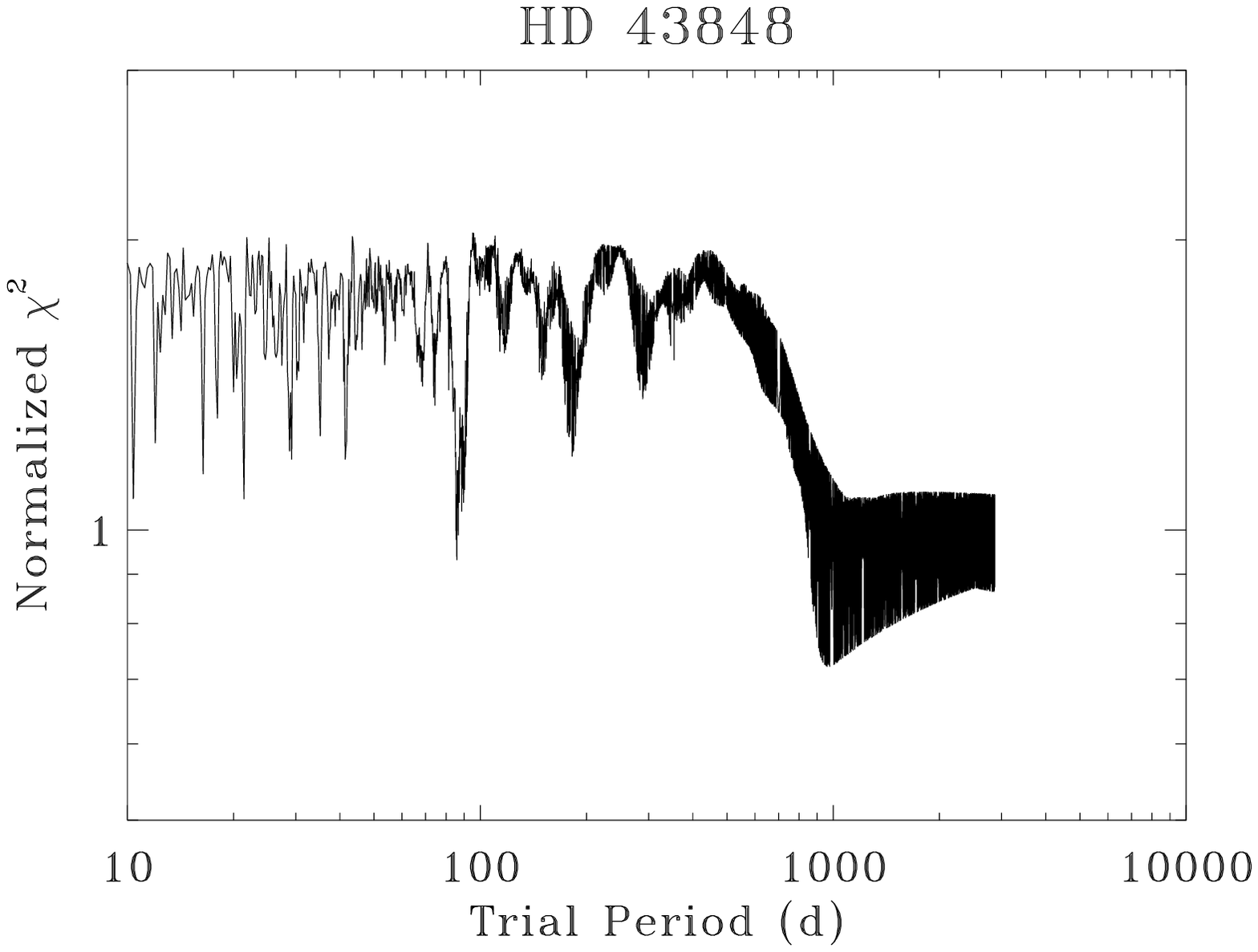} \\
\end{array} $
      \caption{Periodograms of \object{HD 131664} (top) and \object{HD 43848} (bottom) assuming $e$ and $T_0$ from the
known spectroscopic orbit.              }
         \label{per1}
   \end{figure}

As a first step in the process, we decorrelated and weighted all available Hipparcos
along-scan measurements for the two stars, following the prescriptions of~\citet{pourbaix00}
and~\citet{van98}. We then asked whether an orbital model could result in an improved
description of the Hipparcos data, and used the Thiele-Innes representation of a photocentric orbit~\citep{heintz78}
to carry out a linear least squares fit over a large grid of periods bracketing the
ones obtained from the radial velocity measurements, while keeping fixed $e$
and $T_0$ to their spectroscopic values. We thus seeked to minimize
\begin{equation}
\chi^2 = \left(\Delta\nu-\sum_{k=1}^{9}\frac{\partial\nu}{\partial p_k}\Delta p_k\right)^t
         \Sigma^{-1}
         \left(\Delta\nu-\sum_{k=1}^{9}\frac{\partial\nu}{\partial p_k}\Delta p_k\right),
\end{equation}

\noindent
where the superscript $t$ indicates transposed, $\Delta\nu$ is the vector of along-scan Hipparcos IAD residuals
to a single-star, 5-parameter fit, $\partial\nu/\partial p_k$ is the vector of partial derivatives
of the along-scan coordinate with respect to the $k$-th fitted parameter, $\Delta p_k$ the relative correction,
and $\Sigma^{-1}$ is the inverse of the covariance matrix of the solution. The fitted model is fully linear
in 9 parameters, i.e. the five astrometric ones and the four Thiele-Innes constants $A$, $B$, $F$, and $G$. The
results of this period search are shown for both stars in Figure~\ref{per1}. The presence of a long-period
trend in the data appears clear for both \object{HD 131664} and \object{HD 43848}, in clear accord with the choice of an acceleration
solution for both objects at the time of publication of the Hipparcos catalogue.
\begin{table*}[t]
\caption{Orbital fits to the Hipparcos IAD for \object{HD 131664} and \object{HD 43848},
assuming knowledge of $P$, $e$, $T_0$, and $\omega$ from the
spectroscopic orbit. Adjustable parameters were corrections to the
five astrometric parameters, $i$, and $\Omega$. The resulting estimated
masses of the companions are also given, based on the fitted value of the
inclination angle.}
\label{tabres}      
\centering                          
\begin{tabular}{c c c c c c c c c c c c}        
\hline\hline                 
Star Name & $\Delta\alpha$ & $\Delta\delta$ & $\Delta\pi_\star$ & $\Delta\mu_{\alpha^\star}$
& $\Delta\mu_\delta$ & $i$ & $\Omega$ & $M_c$ & $\chi^2_\nu(5)$ & $\chi^2_\nu(7)$ & $P(F)$\\
& (mas) & (mas) & (mas) & (mas yr$^{-1}$) & (mas yr$^{-1}$) & (deg) & (deg) & ($M_J$) & & & \\
\hline
& & & & & & & & & & & \\
HD 131664 & $-2.8\pm0.6$& $-2.1\pm0.7$&$0.5\pm0.9$ &$-5.1\pm0.7$ &$-4.1\pm0.8$ &$55\pm33$ &$22\pm28$ &$23_{-5}^{+26}$ & 1.39& 1.15& 0.001\\
& & & & & & & & & & & \\
\hline
& & & & & & & & & & & \\
HD 43848 & $-4.9\pm0.7$& $-9.8\pm0.8$& $0.1\pm0.9$& $-8.6\pm0.3$& $0.3\pm0.9$& $12\pm7$& $288\pm22$& $120^{+167}_{-43}$& 1.67& 1.22& 0.0002\\
& & & & & & & & & & & \\
\hline
\end{tabular}
\end{table*}
The orbital period is well-known
from spectroscopy in the case of \object{HD 131664}~\citet{moutou09} report uncertainties on the order of $\sim2\%$),
but less so for \object{HD 43848} (uncertainties reported by~\citet{minniti09} are on the order of $\sim 17\%$).
Despite this, and the fact that Hipparcos only covered a fraction of the orbit path of both objects,
it is nonetheless worthwhile to investigate whether a full-fledged orbital solution can be extracted from the
Hipparcos data, which robustly passes statistical screening.
Following a procedure similar to that adopted by others and outlined above, we then fitted the Hipparcos
IAD for the two objects keeping $P$, $e$, and $T_0$ fixed at their spectroscopic values,
while adjusting the five correction terms to the astrometric parameters
\footnote{We used in the fitting process both $\pi_\star$ and its transformed value
$\pi_\star^\prime = \log\pi_\star$, a trick utilized by~\citet{pourbaix00} to prevent
parallaxes from becoming negative, and converting back to $\pi_\star$ at the
end of the minimization procedure. No significant differences in the final results were found.}, $i$ and $\Omega$
(with the additional constraint of Eq.~\ref{constraint}). We adopted a dense two-dimensional grid of starting
values for both $i$ and $\Omega$, in order to solve a local non-linear minimization problem in which
we used a model function expressed in terms of the Campbell orbital elements. In order to assess the
impact of the more or less precise knowledge of the spectroscopic orbit, the procedure was re-run 100
times, each with a different set of (fixed) spectroscopic orbital elements drawn from gaussian distributions with
mean and standard deviation the best-fit value and its formal error, respectively.

   \begin{figure}
   \centering
   \includegraphics[width=0.5\textwidth]{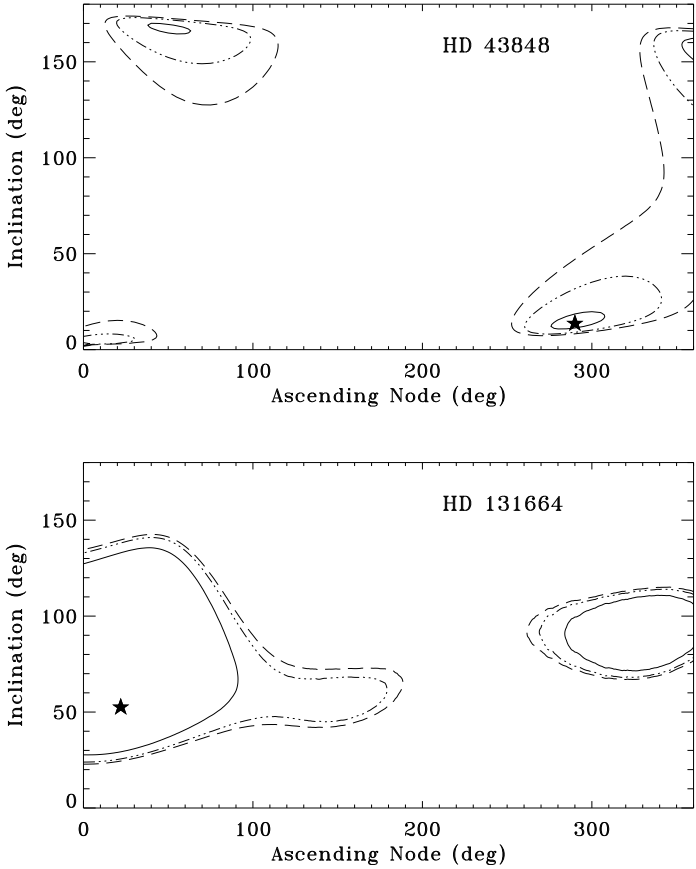}
      \caption{Iso-$\chi^2$ contours corresponding to confidence
      intervals in two dimensions containing 68.3\% (solid line), 90\% (dashed-dotted line),
      and 95\% (dashed line) of the values of $i$ and $\Omega$ as obtained in a 7-parameter fit
      to the Hipparcos IAD of \object{HD 43848} (top) and \object{HD 131664} (bottom), assuming $P$, $e$, $T_0$, and $\omega$
      from the known spectroscopic orbit. The best-fit ($i$,$\Omega$) solutions for both stars
      are indicated with stars.}
         \label{i_om_map}
   \end{figure}

The results of the orbital fit to the Hipparcos IAD of both \object{HD 131664} and \object{HD 43848} are summarized in Table~\ref{tabres}.
The reported uncertainties on $i$ and $\Omega$, as well as the derived mass values for the
companions, take into account the 1-$\sigma$ formal errors on the parameters of the
spectroscopic orbits. The inferred true masses for \object{HD 131664}b and \object{HD 43848}b
are $23_{-5}^{+26}$ $M_J$ and $120^{+167}_{-43}$ $M_J$, respectively. Taken at face value, these numbers would imply that
the companion to \object{HD 131664} has a mass in the brown dwarf regime at the $2-\sigma$ confidence level,
while the unseen object around \object{HD 43848} is likely a low-mass M dwarf (at the $1-\sigma$ confidence level).
An attempt at using the FAST and NDAC data separately for both stars resulted in solutions formally compatible with
the ones reported here, albeit with larger uncertainties and looser constraints on the derived mass estimates.

\begin{figure}
\includegraphics[width=0.45\textwidth]{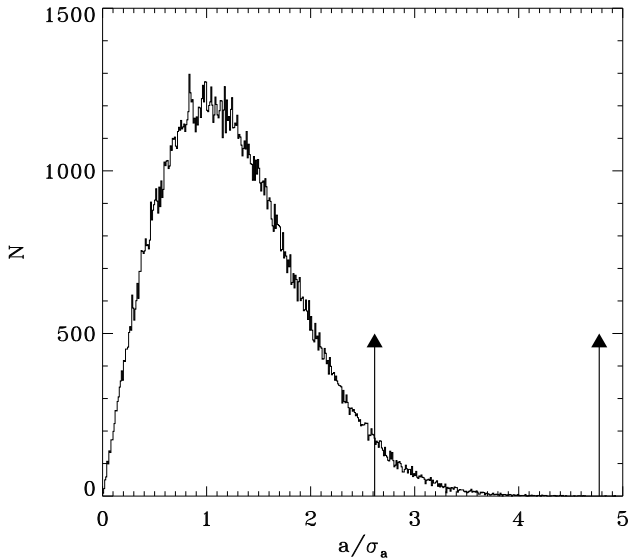}
\caption{Distribution of the ratio $a/\sigma_a$ of false astrometric orbits
for single stars observed by Hipparcos, approximated by a Rayleigh-Rice law following~\citet{halb00}.
The two arrows indicate the values of $a/\sigma_a$ obtained
for \object{HD 43848} and \object{HD 131664} based on a Monte Carlo bootstrap simulation.}
\label{raydist}
\end{figure}

In order to assess the statistical soundness of the derived
astrometric orbits, several checks can be performed. First, based
on an F-test of the null hypothesis that there is no companion, we
find that the addition of two parameters to the model describing
the Hipparcos IAD significantly improves the fit in both cases:
$P(F) = 0.0002$ and $P(F) = 0.001$ for HD 131664 and \object{HD
43848}, respectively. Second, similarly to~\citet{reffert06}, we
have further explored the reliability of the orbital solutions by
inspecting the joint confidence regions (corresponding to
iso-$\chi^2$ contours) in the $i-\Omega$ plane, for a given
statistical level of confidence. As shown in
Figure~\ref{i_om_map}, the 95\% (2-$\sigma$) iso-$\chi^2$ contours
cover relatively narrow regions in the $i-\Omega$ plane, ruling
out with confidence very small inclinations in the case of
\object{HD 131664}, and close to edge-on configurations in the
case of \object{HD 43848}. Ambiguities in the orbit orientation
are clearly seen in both cases, highlighted by local minima in the
$\chi^2$ surface corresponding to configurations with the opposite
sense of revolution. Nevertheless, one would conclude that
Hipparcos astrometry has successfully detected orbital motion
induced by the two massive companions to \object{HD 131664} and
\object{HD 43848}. We consider the 95\% confidence level as the
appropriate measure of the degree of statistical robustness with
which these results can be regarded. Finally, we have applied a
standard bootstrap method~\citep{efron82}, based on 1000 Monte
Carlo resamples with replacement, to the available data for both
stars in order to derive empirical error estimates on the derived
orbital semi-major axes (constrained by Eq.~\ref{constraint}
during the fits). Our procedure closely follows the one adopted
by~\citet{zuck01}, who first realized, due to the assigned
correlation between pairs of Hipparcos measurements for which both
FAST and NDAC data are available, the need to preserve the pairing
of the measurements while drawing new random datasets, in order to
make the bootstrap method applicable in the first place. The two
resulting ratios $a/\sigma_a$ have then been compared to the
Rayleigh-Rice law distribution of {\it false} $a/\sigma_{a}$ one
can expect to obtain in the case of Hipparcos data of single stars
(see~\citep{halb00}). As shown in Figure~\ref{raydist}, the
nominal values of $a/\sigma_a$ for \object{HD 131664} and
\object{HD 43848} fall in the tail of the Rayleigh-Rice
distribution: the probability that $a/\sigma_a$ exceeds the
observed one is found to be 0.03 and $2\times10^{-5}$ for
\object{HD 43848} and \object{HD 131664}, respectively. This can
be again interpreted as a detection of a significant Hipparcos
astrometric orbit for both stars, at the 97\% and 99.99\%
confidence level, respectively. For the purpose of this study, we
adopt the more conservative abovementioned 95\% confidence level
to gauge the actual degree of statistical confidence with which
the orbits of \object{HD 43848}b and HD13664b are considered
detected.

\section{Summary, discussion, and conclusions}\label{discuss}

We have inspected the Hipparcos IAD for two stars, \object{HD 43848} and \object{HD 131664}, with Doppler-detected
companions with minimum masses in the brown dwarf regime. We have presented a body of
supporting evidence which appears to confirm the detection of orbital motion in the Hipparcos IAD
of \object{HD 43848} and \object{HD 131664}, at a somewhat significant (95\%) level of statistical confidence.
The inferred actual masses of \object{HD 43848}b and \object{HD 131664}b are found to be $M_c = 120^{+167}_{-43}$ $M_J$
and $M_c = 23^{+26}_{-5}$ $M_J$, respectively. The former thus
appears to be a late M dwarf (at the $\sim1-\sigma$ confidence level),
while the latter appears to be a brown dwarf (at the $\sim2-\sigma$ confidence level).
Taken at face value, \object{HD 131664b} is nominally the lowest-mass brown dwarf confirmed with
a combination of Hipparcos and precision Doppler measurements (the astrometric orbits of giant
planets obtained with HST/FGS + RV data lie in a different ballpark).
The larger uncertainties in the spectroscopic orbital elements of \object{HD 43848}, combined with the low
value of $i$ inferred from the orbital fit, translate in larger uncertainties
in the mass estimate, leaving margin for this unseen object to also be a brown dwarf (as well as a higher-mass
M dwarf).

Based on the mass-luminosity relations of \cite{delfosse},
\object{HD 43848b}, with its nominal mass of 120 $M_J$, is expected to have magnitude differences
of about 8.7, 5.7, 5.4, and 5.2 in the {\em V, J, H}, and {\em K} bands, respectively.
With a projected separation of about 0.15 arcsec at apoastron
it should be detectable with current AO instruments at 8m class telescopes.
The magnitude difference between \object{HD 131664b} and its parent star depends critically on
the age of the system. \cite{moutou09} give an age of $2.4\pm1.8$ Gyr for \object{HD 131664} based on isochrone fitting.
Additional clues can be derived from stellar activity indicators.
\cite{moutou09} measured $\log R^{'}_{HK}=-4.82\pm0.07$. This corresponds to an age of 3.5 Gyr using
the calibration by \cite{mamajek08}. We also searched for X-ray emission from the system.
The ROSAT Faint Source Catalog~\citep{rosatfaint} includes a source (\object{1RXS J150003.3-733128})
at 40 arcsec from \object{HD 131664} (with a quoted positional error of 29 arcsec). The association is then
doubtful. Assuming \object{1RXS J150003.3-733128} is the X-ray counterpart of \object{HD 131664}, we
derived (using the flux calibration by \cite{hunsch}) $\log L_X = 28.5$ and an age of 2.1 Gyr
(using the calibration by \cite{mamajek08}). Alternatively, a null detection would imply
an older age. Stellar activity indicators then exclude the youngest stellar ages compatible with
isochrone fitting. A more plausible lower limit to the stellar age is about 1.5 Gyr.
\object{HD 131664b}, given our best-fit mass value and the age  of 2.4 Gyr, is expected to have
magnitude differences of about 15.0, 15.2, and 17.2
in {\em J, H,} and {\em K} bands respectively, based on the models of \cite{baraffe03}.
The corresponding effective temperature is about 700 K, at the cool end of the currently known T dwarfs.
Considering the small projected separation ($<0.1$ arcsec), such a contrast is not achievable with
current instrumentation, and it would also be challenging for the next generation of direct imaging
instruments.

Any successful attempt, such as the one presented here, at deriving true masses of the substellar companion candidates
detected from radial velocity surveys is definitely worthwhile, as the results can be seen in the
context of the observed paucity of close brown dwarf companions to solar-type stars in the solar neighborhood,
the well-known brown dwarf desert, and in particular their binarity properties.
For example, taken at face value the rather large eccentricity of
\object{HD 131664b} ($e=0.638$) supports the notion that massive planetary companions
and brown dwarfs are preferably found on eccentric orbits \citep{ribas07}.
On the other hand, the high metallicity of the parent star ([Fe/H=+0.32) does not follow the proposed trend of
metallicity vs. mass for substellar companions \citep{ribas07}. While not conclusive,
these findings bring new/updated information that can certainly help to better our understanding of the
formation mechanism of such objects.

\begin{table*}[h]
\begin{minipage}[t]{1.0\textwidth}   
\caption{Close companions to solar-type stars with (projected) masses between 10 $M_J$ and 80 $M_J$.
The first part lists the probable true substellar objects,
the second part the objects for which only the projected masses or poor constraints on the true mass are available, the third part the
companions with true masses above the substellar limit. In the remarks, PL refers to objects with additional companions of planetary mass,
MULT multiple systems (additional stellar companions), HIPG stars with Hipparcos acceleration solutions.}             
\label{t:bd}      
\centering                          
\renewcommand{\footnoterule}{}  

{\scriptsize
\begin{tabular}{c c c c c c c c c c l}        
\hline\hline                 
Object & $M_c\sin i$ & Period & a & e & i & $M_c$ & SpT & Mstar & [Fe/H] & Remarks and References \\    
       & $ M_{J}$  & d      & AU&   & deg & $ M_{J}$  & & $M_{\odot}$ & \\
\hline                        

\hline
\multicolumn{11}{c}{Probable substellar objects with mass $M_c > 10 M_{J}$ and $a<7$ AU around solar-type stars} \\
\hline

\object{HD29587b}      &  41    & 1474.9   &       &   0.356  &             & $41~(27)$        &  G2V  & 1.00  &  -0.61  &   \cite{halb00}, \cite{bensby2005}
        \footnote{For spectroscopic binaries with brown dwarf candidates observed by Hipparcos and with an actual
mass estimate derived by~\citet{halb00}, the error distribution of the secondary masses is not Gaussian and not
symmetric. We have listed here the values of $M_c$ and the accompanying one-sided errors as defined by Eq.~5 in~\citet{halb00}.
The star has both chemical abundances and kinematics of thick disk. } \\
\object{HD38529c}      &  12.7  &  2174.3  & 3.68  &   0.36   & 160         & $37^{+36}_{-19}$ & G4IV  & 1.46  &  +0.45  & \cite{butler06}, \cite{reffert06}, \\
&  & & & & & & & & & \cite{vf}, PL, MULT, HIPG  \\  
\object{HD89707b}      &  59    & 297.708  &       &   0.952  &             & $64~(19)$        &  G1V  & 1.05  &  -0.42  &  \cite{halb00}, \cite{edvarssson93}
\footnote{Thick disk population.} \\
\object{HD127506b}     &  36    &  2599    &       &   0.716  &             & $45~(21)$        &  K3V  & 0.75  &  +0.06  &  \cite{halb00}, \cite{haywood01}  \\
\object{HD131664b}     &  18.15 &  1951    &  3.17 &   0.638  & $55\pm12$   & $23^{+26}_{-5}$   & G3V   & 1.10  &  +0.32  & this paper, HIPG \\
\object{HD168443c}     &  18.1  &  1765.8  & 2.91  &   0.2125 & 150         & $34\pm12$        & G5    & 1.08  &  +0.08  & \cite{reffert06}, PL \\ 
              &        &          &       &          &             & $<82$            &       &       &         & \cite{zuck01} \\
\object{CoRoT-3b}      &        &  4.2568  & 0.057 &    0.0   & 85.9        & $21.7\pm1.0$     & F3V   & 1.37  &  -0.02  & \cite{deleuil08}, transit
\footnote{Transiting system.~\citet{triaud09} found a significant misalignment between the planetary orbital axis and
                       the stellar rotation axis ($37.6^{+10.0}_{-22.3}$ deg).} \\
\object{XO-3b}         &        &  3.19152 & 0.0454&   0.26   & 84.20       & $11.79\pm0.59$   & F5V   & 1.21  &  -0.18  & \cite{winn08},  transit
             \footnote{Transiting system.~\citet{winn09} found a significant misalignment between the planetary orbital axis and
                       the stellar rotation axis ($37.3\pm3.7$ deg).} \\
\hline
\multicolumn{11}{c}{Objects with projected mass $10 < M_c \sin i< 80 $ $M_J$ and $a<7$ AU around solar-type stars} \\
\hline
\object{HD4747b}       &  42.3  & 6832.0   & 6.70  &   0.64   &       &           &       & 0.82  & -0.22   &  \cite{nidever02}, \cite{vf} \\  
\object{HD13507b}      &  52    & 3000     & 4.3   &   0.14   &       &           &  G0   & 1.00  & +0.03   &  \cite{perrier}, \cite{vf} 
          \footnote{The orbit in \citet{perrier} is preliminary, with only limited phase coverage.~\citet{perrier} also reported no detection with AO.}\\
\object{HD16760b}      &  13.13 & 466.47   & 1.084 &   0.084  &       &           &  G5V  & 0.78 & +0.07  & \cite{sato09}, \cite{bouchy09}, MULT  \\
\object{HD30339b}      &  77.8  & 15.0778  & 0.13  &   0.25   &       &           &  F8   & 1.39 & +0.26  & \cite{nidever02}, \cite{vf} \\  
\object{HD39091b}      &  10.35 &  2063.818& 3.29  &   0.62   &       &           &  G1IV & 1.10  &  +0.05  &  \cite{butler06},  \cite{vf} \\ 
\object{HD65430b}      &  67.8  & 3138.0   & 4.00  &   0.32   &       &           &       & 0.83  &  -0.12  &  \cite{nidever02}, \cite{vf} \\ 
\object{HD91669b}      &  30.6  &  497.5   & 1.205 &   0.448  &       &           &       & 0.91  &  +0.31  &  \cite{witten09},  \\
\object{HD98230b}      &  35    &  3.9805  &       &   0.0    &       &           &       & 0.86  &  -0.35  & \cite{fuhrmann08}, MULT
     \footnote{Assuming synchornous rotation and orbit aligment~\citet{fuhrmann08} derived $i=15$ and $M_c=0.14_{-0.05}^{+0.09}~M_{\odot}$
               for the companion. The system is a quadruple with another SB at $a=2.53$ arcsec (21 AU) with e=0.41  from \object{HD98230}.} \\
\object{HD114762b}      &  11.02 &  83.90   & 0.3   &   0.34   &             &                &  F9V  & 0.93  &  -0.65  & \cite{halb00}, MULT  
\footnote{The possibility that the star is seen nearly pole-on because of its very small projected rotational velocity is widely
           debated in the literature. \cite{halb00} derived a very uncertain astrometric mass $112~M_J$ with an error of
           $103~M_{J}$. Additional low-mass companion close to the substellar boundary detected at 130 AU by
           \cite{patience02}. Member of the thick disk population.}\\
\object{HD136118b}      &  11.9  &  1209    & 2.3   &   0.37   &       &           &  F9V  & 1.25  &  -0.05  &  \cite{butler06}, \cite{vf} \\ 
\object{HD137510}       &  26    &  798.2   & 1.85  &   0.402  & $>16$ & $<94$     &  G0IV & 1.41   & +0.37  &  \cite{endl04}, \cite{vf}  \\ 
\object{HD156846b}      &  10.45 &  359.51  & 0.99  &   0.8472 &       &           &  G0V  & 1.43  &  +0.22  &  \cite{tamuz08}, MULT   \\ 
\object{HD162020b}      &  13.75 & 8.428198 & 0.072 &   0.277  &       &           &  K2V  & 0.78  &  +0.11  &  \cite{udry02}
          \footnote{\citet{zuck01} derived an astromeric solution that gives a companion mass of $3.0\pm0.94~M_{\odot}$. Not compatible with
           the stellar properties of the primary. Probably spurious due to the short orbital period.} \\
\object{HD167665b}      &  50.3  & 4385     & 5.47  &   0.337  &       &           &  G0V  & 1.11  & -0.17   & \cite{patel07}, \cite{vf} \\  
\object{HD174457b}      &  65.8  & 840.800  & 1.90  &   0.23   &       &           &       & 1.07  & -0.18   & \cite{nidever02}, \cite{vf} \\  
\object{HD184860b}      &  32.0  &  693     & 1.4   &   0.67   &       &           &  K2V  & 0.77  &  -0.04  & \cite{vogt02}, \cite{vf}, MULT \\ 

\object{HD191760b}      &  38.17 &  505.65  & 1.35  &   0.63   &       &           & G3IV/V & 1.28 & +0.29 &   \cite{jenkins09} \\
\object{HD202206b}      &  17.4  &  255.87  & 0.83  &   0.435  &       & $<149$    &  G6V  & 1.17  & +0.35  &  \cite{correia05}, \cite{zuck01}, PL \\ 
\object{HD283750b}      &   51   &  1.788   &       & 0.002    &       &           &  dK5  & 0.67  & +0.03   & \cite{halb00}, MULT
            \footnote{\cite{halb00} derived  an astrometric mass  of $182~M_{J}$ with a large error of $470~M_{J}$.
             \cite{glebocki95} estimate $i=22\pm10$ and then $M=136\pm39~M_{J}$ from the rotation period (synchronous with the orbit) and projected
             rotational velocity, assuming the rotation axis is perpendicular to the orbital plane. However, misalignements can not be ruled out,
             especially considering that the system has a wide WD companion. The system is probably associated with the Hyades \citep{catalan08}.
             Metallicity from \cite{catalan08}.} \\
\object{TYC 2534-698-1b}&  39.7  & 103.698  & 0.44  &   0.385  &       &           &  G2V  & 1.00 & -0.25  & \cite{kane09}  \\
\object{HAT-P-13c}      &  15.2  & 428.5    & 1.19  &   0.691  &       &           &  G4V  & 1.22 & +0.43  & \cite{bakos09}, PL  \\

\hline
\multicolumn{11}{c}{Probable stellar companions with projected mass in the substellar range} \\
\hline
\object{HD18445B}       &  45    &  554.6   &       &    0.558 &             & $187~(20)$          &  K2V  & 0.78  &   0.00  & \cite{halb00}, \cite{vf}, MULT   
     \footnote{Also resolved with direct imaging by \cite{beuzit04}. Quintuple system (see \cite{bd07} and references therein). } \\
               &        &          &       &          & -16.1       & $178\pm20$          &       &       &         & \cite{zuck01} \\
\object{HD33636B}       &  9.3   &  2128    &       &    0.48  & $4.1\pm0.1$ & $142^{+3.3}_{-1.8}$ &       & 1.02  &  -0.13  & \cite{bean07}\\ 
\object{HD43848B}       &  25    &  2371    & 3.4   &   0.69   &  $55\pm12$  & $120^{+167}_{-43}$  & G0V   & 0.89  &  -0.03  & this paper, HIPG \\
\object{HD110833B}      &  17    &  271.17  &       &   0.784  &             & $146~(12)$          & K3V   & 0.72  &   0.00  & \cite{halb00}, \cite{mishenina04}
       \footnote{Purely astrometric orbital solution in Hipparcos.}  \\ 
               &        &          &       &          & 7.76        & $143\pm12$          &       &       &         & \cite{zuck01} \\
\object{HD112758B}      &  34    &  103.258 &       &   0.139  &             & $212~(44)$          & K0V   & 0.79  &  -0.56  & \cite{halb00}, \cite{mishenina04} \\ 
               &        &          &       &          & 10.4        & $213\pm43$          &       &       &         & \cite{zuck01} \\
\object{HD140913B}      &  43.2  & 147.968  & 0.55  &   0.54   &             &  $177~(73)$         &       & 1.17  & +0.13   & \cite{nidever02}, \cite{halb00}  \\  
               &        &          &       &          &  16.3       &  $181\pm74$         &       &       &         & \cite{zuck01}, \cite{vf}
                   \footnote{The significance of the orbital solution in \cite{zuck01} is about 0.92.} \\
\object{HD164427B}      &  46    &  108.55  &  0.46 &    0.55  & 8.5         & $372^{+106}_{-85}$  &       & 1.18  & +0.13   & \cite{zuck01} \\
               &        &          &       &          &             & $<190$              &       &       &         & \cite{tinney01}  \\
\object{HD169822B}      &  27.2  &  293.1   &  0.84 &    0.48  & 175         & $320$               &  G5V  & 0.91  & -0.12   & \cite{vogt02}, MULT  
       \footnote{Triple system, see \cite{bd07}. Flagged as stochastic solution in Hipparcos.} \\
\object{HD217580B}      &  68    &  454.66  &       &  0.520   &             & $171~(14)$          &  K4V  & 0.69  &  -0.06  & \cite{halb00}, \cite{gray03}
                      \footnote{Purely astrometric orbital solution in Hipparcos.} \\ %
               &        &          &       &          & 25.2        & $172\pm15$          &       &       &         & \cite{zuck01}  \\
\object{BD-04 782B}     &  48    &  716.68  &       &    0.074 &             & $261~(15)$          &  K5V  &  0.67 &  +0.05  & \cite{halb00}, \cite{heiter03}
   \footnote{Flagged as stochastic solution in Hipparcos.}  \\ 
               &        &          &       &          & 12.77       & $257\pm16$          &       &       &         & \cite{zuck01} \\

\hline                                   
\end{tabular}
}
\end{minipage}
\end{table*}

\subsection{Properties of massive planets and brown dwarf companions to solar-type stars}

To put our discussion in a more general context, we summarize in
Table~\ref{t:bd} the main properties of the companions with
projected masses between 10 and 80 $M_J$, and semimajor axis
smaller than $\approx 7$ AU, orbiting main-sequence stars with
masses between 0.7 to 1.5 $M_{\odot}$. This selection matches the
sample of most high-precision radial velocity surveys and avoids
the additional complications of large variations in the stellar
mass and evolutionary status (very small number statistics being
one of the most relevant). The literature data collected here are
meant to provide ease of consultation and reference for future
works on the many outstanding issues we touch upon thereafter. In
the Table, the determination of or constraints on the inclination
and true companion mass derived from astrometry or transit
photometry are included when available. For completeness and
reference, we also include companions with substellar projected
mass and astrometrically derived stellar mass. We have also
included \object{HD 33636}, whose companion with $M_c \sin i$ just
below the adopted threshold was shown to be a low mass star
\citep{bean07}. Orbital elements and projected masses are from
listed discovery papers,~\citet{butler06} or~\cite{halb00}.
Stellar masses and metallicities are from~\cite{vf}, discovery
papers or additional references. Further information, such as
additional stellar or planetary companions, the sources for the
metallicity values, and the specific treatment of some of the mass
and error estimates, can be found in the notes.

\begin{figure}
\centering
$\begin{array}{c}
\includegraphics[width=0.45\textwidth]{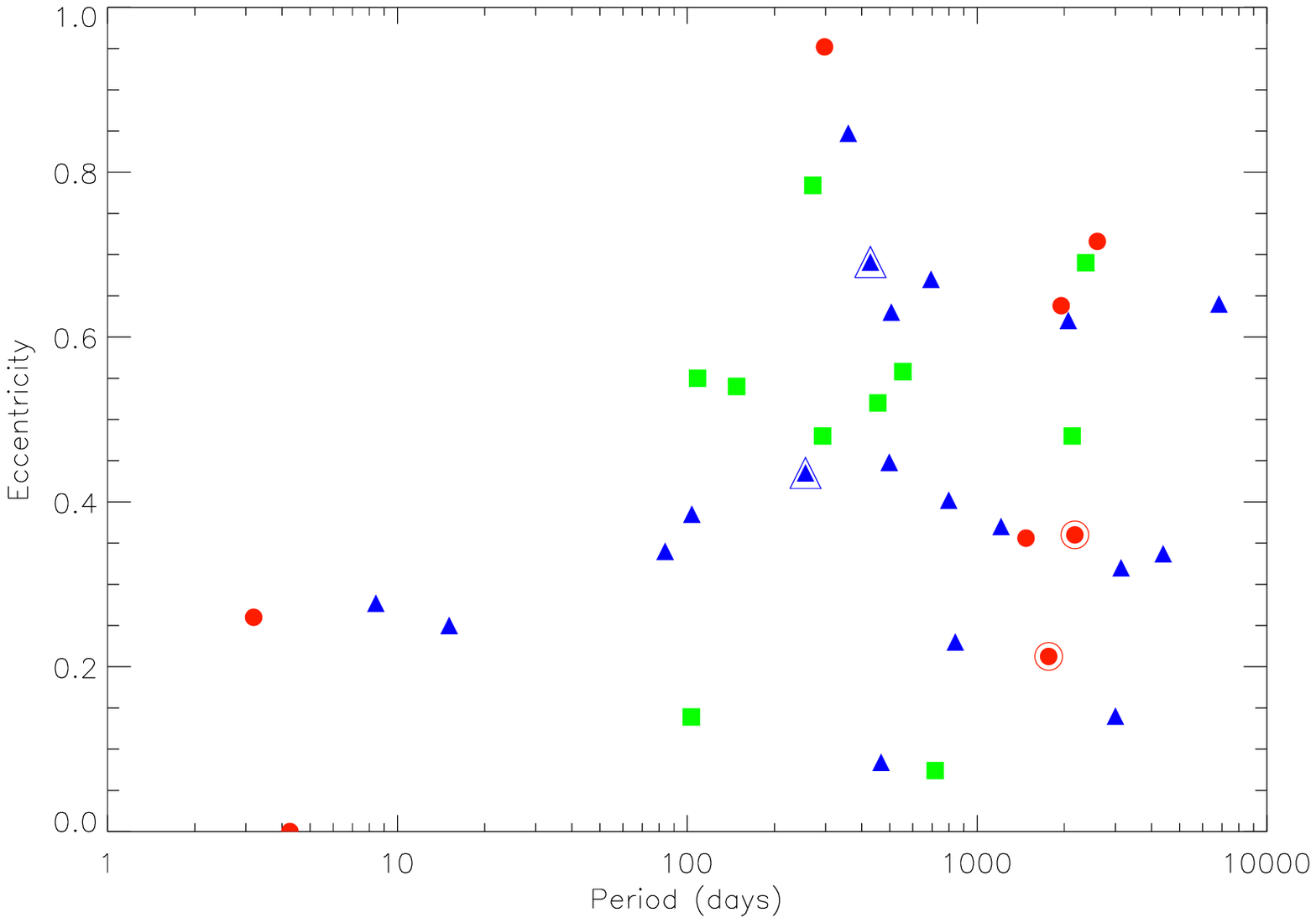}\\
\includegraphics[width=0.45\textwidth]{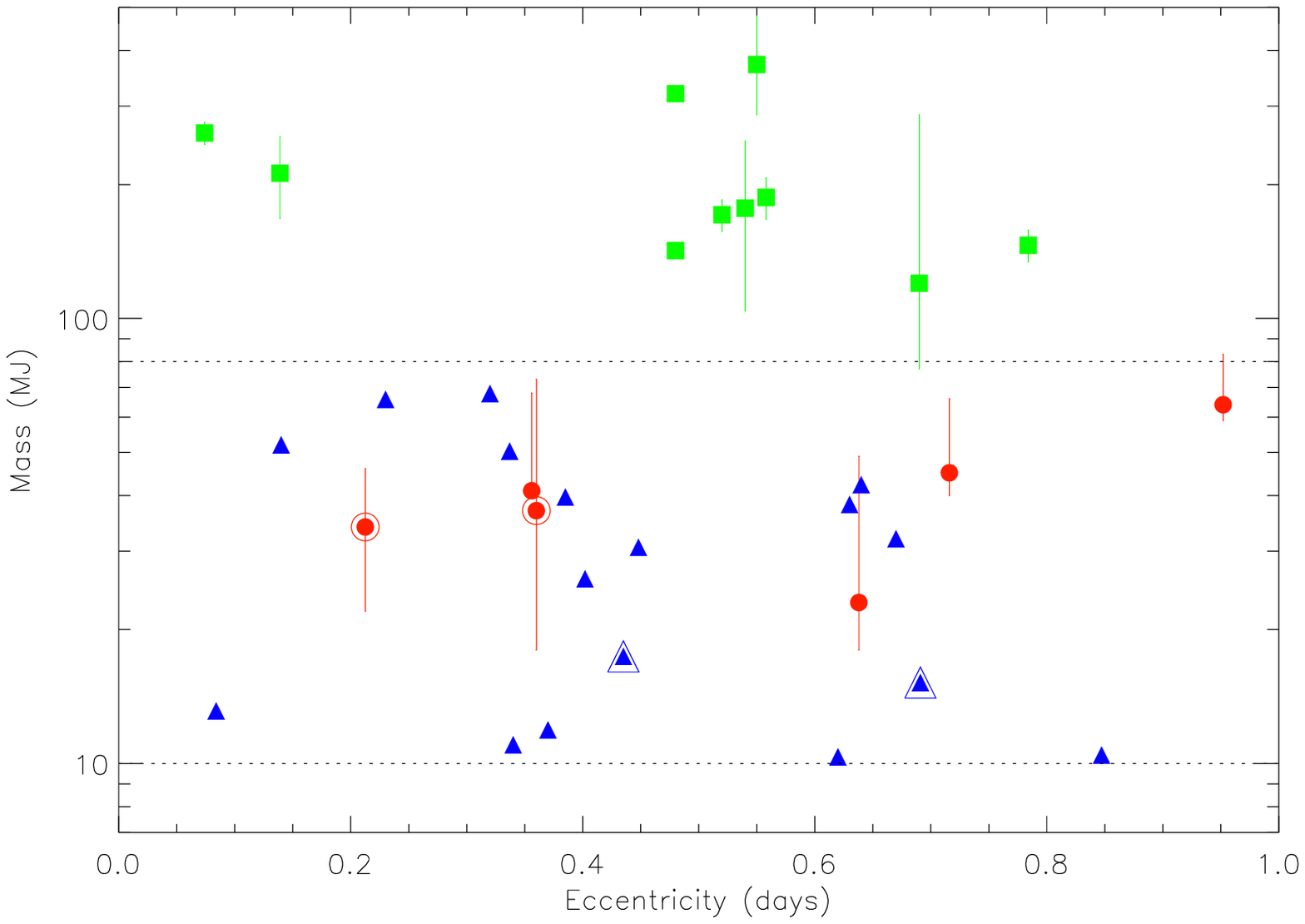}\\
\end{array}$
\caption{Period vs eccentricity (top panel) and eccentricity vs companion mass
(bottom panel). In the lower panel plot only objects with period longer than 20 days are shown).
Red filled circles: companions with true estimated masses in the range 10-80 $M_J$. Blue
triangles: companions with projected masses in the range 10-80 $M_J$. Green
squares: companions with projected masses in the range $<80 M_J$ but true stellar
masses $> 80 M_J$. The three objects with larger symbols belong to the
multiple systems \object{HD 38529}, \object{HD 168443} and \object{HD 202206}.}
\label{f:plot1}
\end{figure}

\begin{figure}
\centering
$\begin{array}{c}
\includegraphics[width=0.45\textwidth]{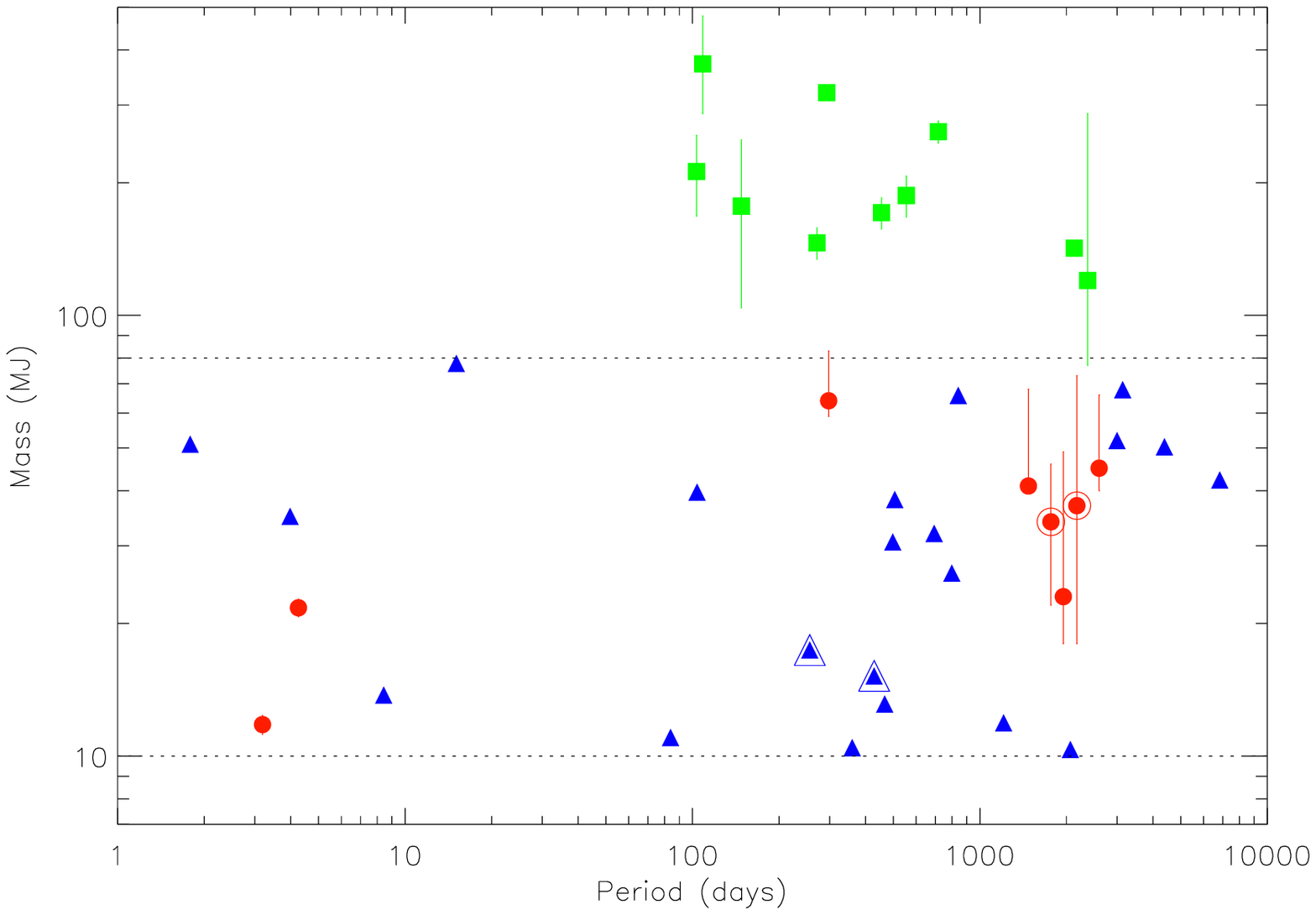}\\
\includegraphics[width=0.45\textwidth]{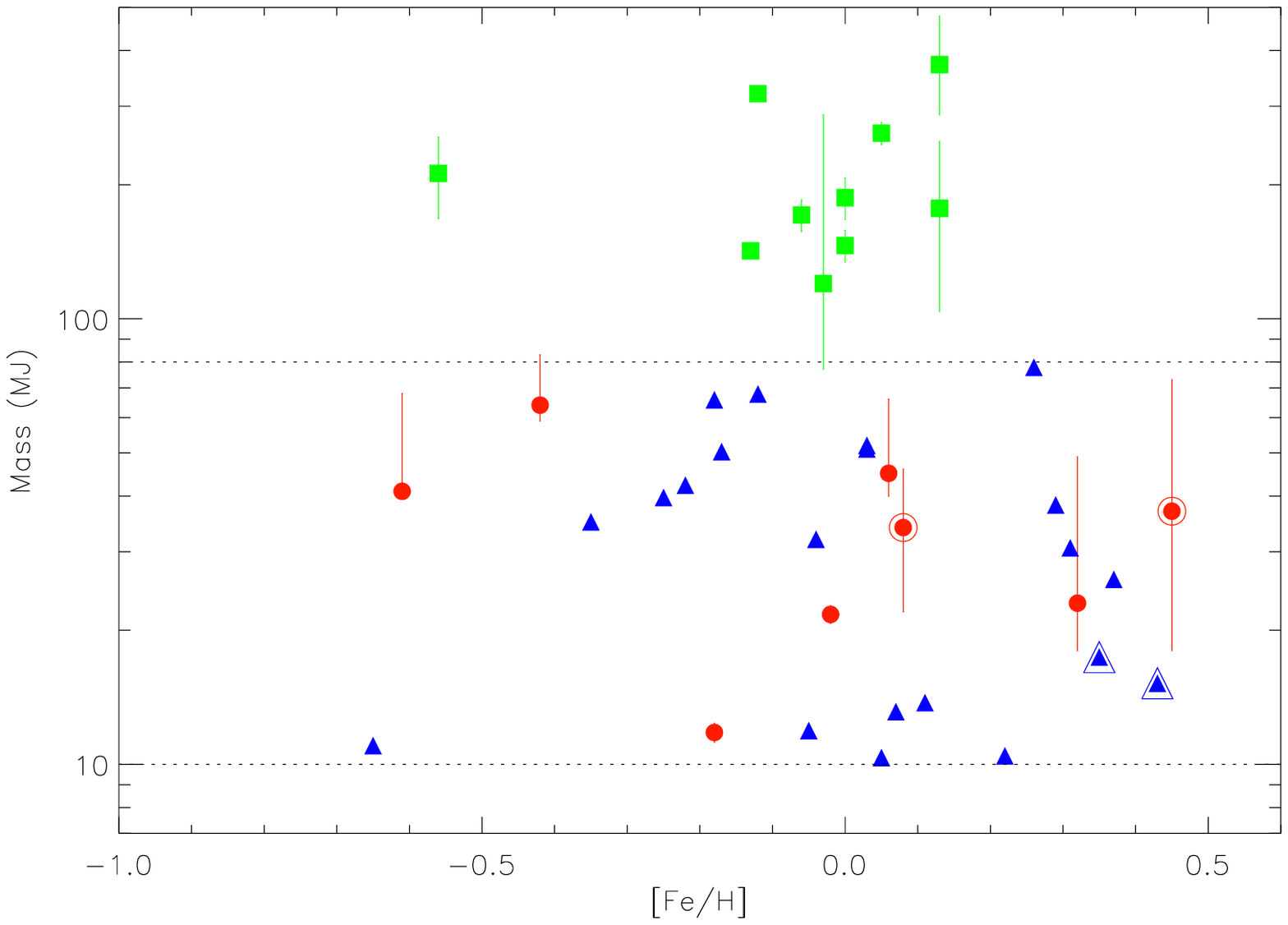}\\
\includegraphics[width=0.45\textwidth]{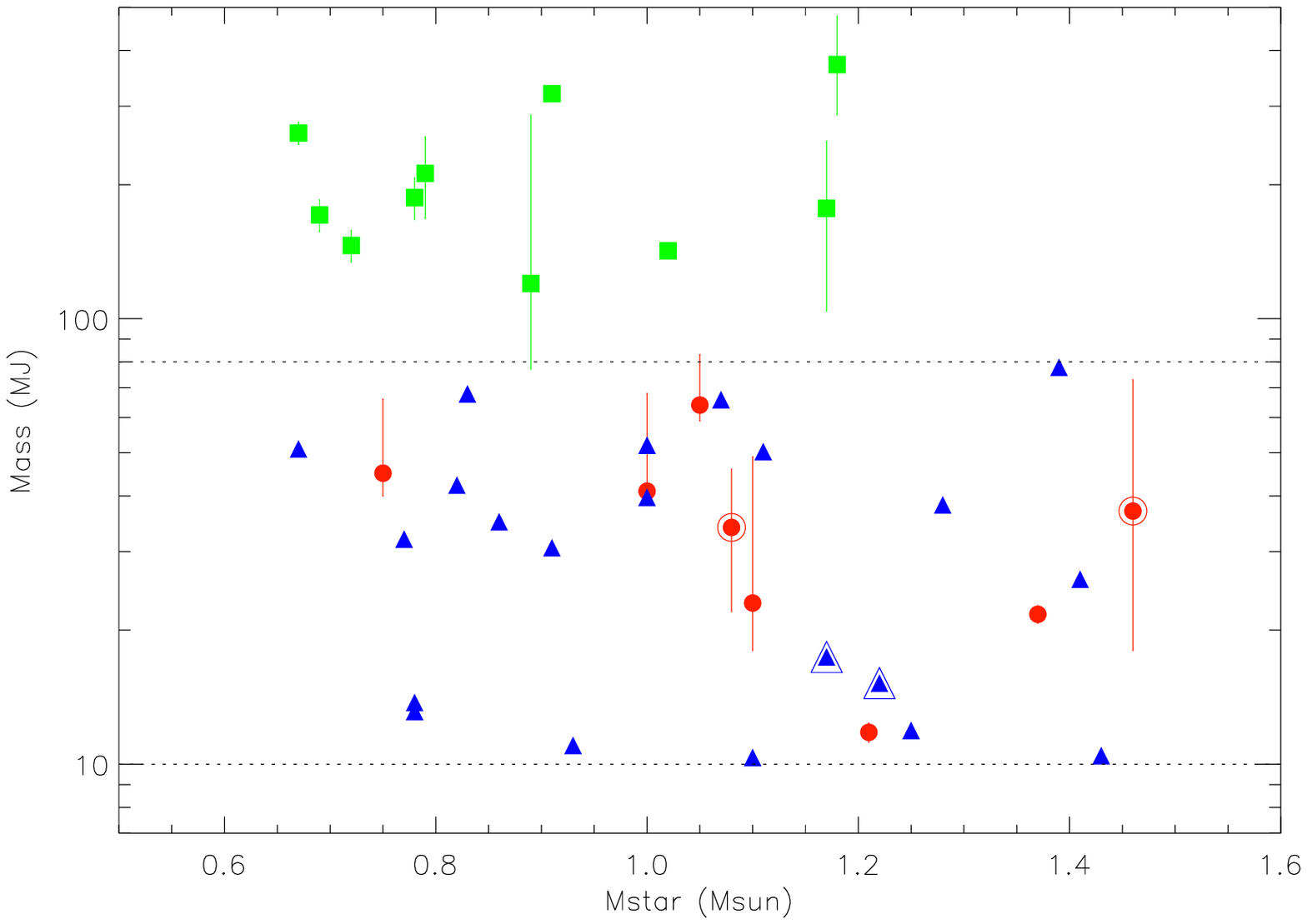}\\
\end{array}$
\caption{Orbital period  vs companion mass (top), metallicity
vs companion mass (center), stellar mass vs companion mass (bottom).
Symbols as in the previous Figure.}
\label{f:plot2}
\end{figure}

As discussed in the Introduction,~\citet{halb00} have shown how a
significant fraction of Doppler-detected candidate substellar
companions were in fact low-mass stars viewed at low inclination.
Nevertheless, a few candidates have masses firmly in the
substellar regime, partially filling the brown dwarf desert.
Interestingly, in two cases (\object{HD 38529} and \object{HD
168443}) additional companions in the planetary mass regime were
found. \footnote{Two additional probable cases of a system with a
brown dwarf and a planet in close orbits are those of
\object{HAT-P-13} \citep{bakos09} and \object{HD 202206}
\citep{correia05}. In the first case, the planetary nature of the
inner companion is confirmed by the occurrence of the transit. The
true masses of the more massive companions are not available but
there are perspectives for deriving them thanks to the strong
dynamical interactions in the case of \object{HD 202206} and the
transit timing variations in the case of \object{HAT-P-13}.} We
note that such system configurations are more typical of planetary
systems than of multiple stellar systems.

In Figure~\ref{f:plot1} and Figure~\ref{f:plot2} we show a set of correlation diagrams among the most relevant
quantities of the systems collected in Table~\ref{t:bd}.

\begin{figure}
\centering
$\begin{array}{c}
\includegraphics[width=0.45\textwidth]{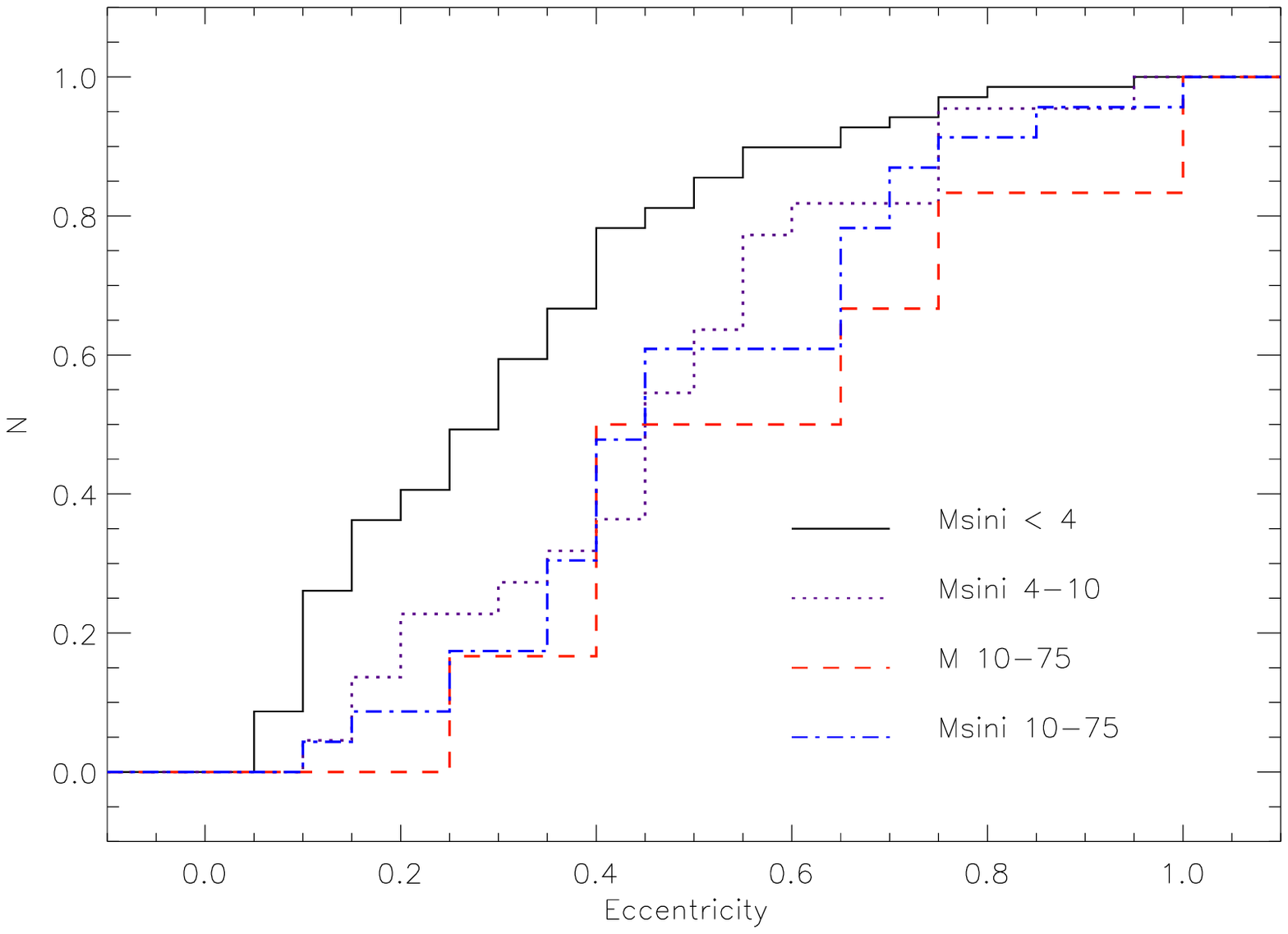}\\
\includegraphics[width=0.45\textwidth]{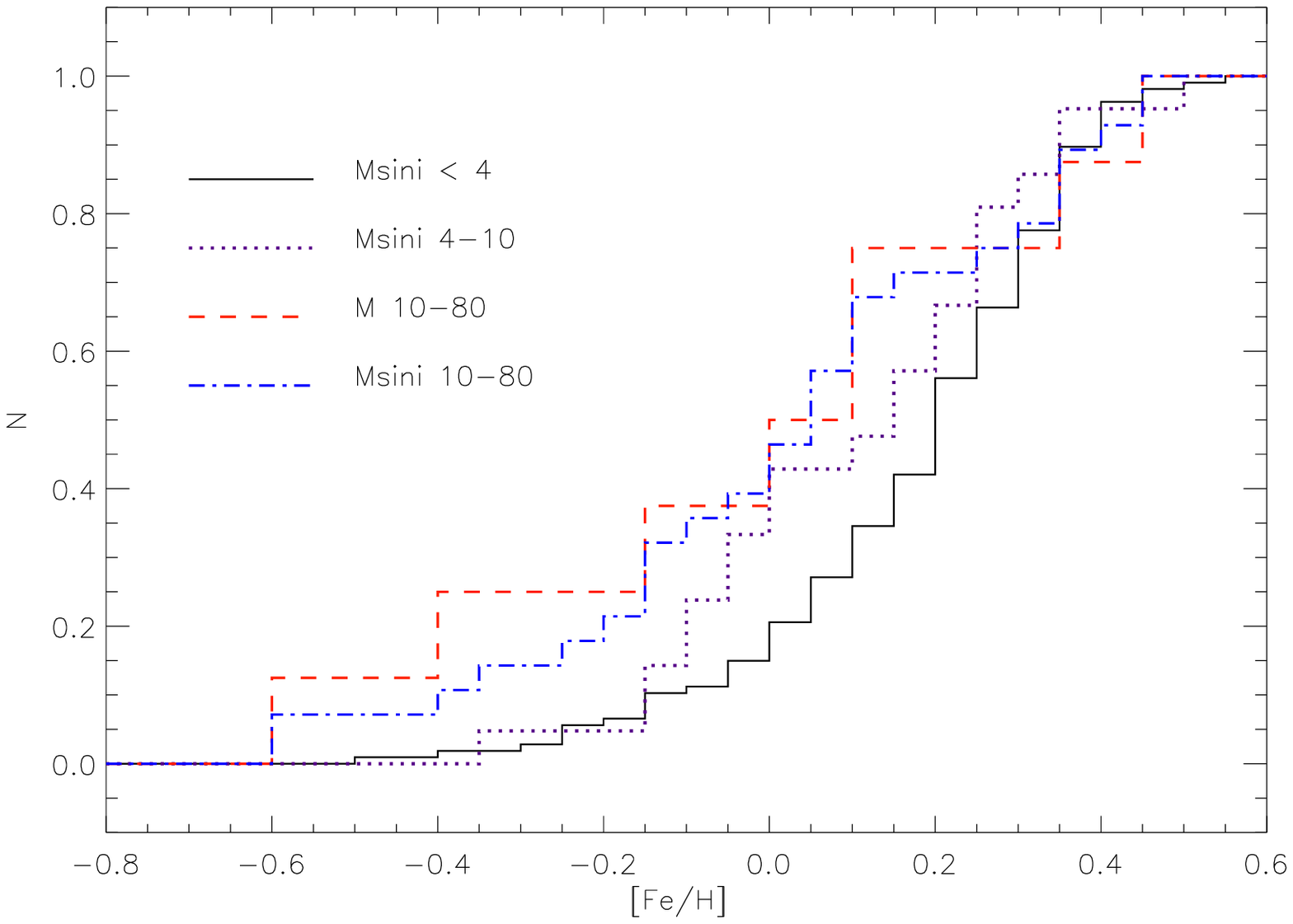}\\
\end{array}$
\caption{Upper panel: cumulative distribution of the eccentricity of objects with period longer than 20 days
orbiting main-sequence stars with mass between 0.7 and 1.5 $M_{\odot}$. Continuous (black) line:
planets with $M_c \sin i < 4~M_J$ (from \cite{butler06}). Dotted (purple) line:
planets with $4 < M_c \sin i < 10~M_J$ (from \cite{butler06}).
Dashed (red) line: objects with mass $10 < M_c < 80~M_J$ (this paper),
Dashed-dot (blue) line: objects with  $10 < M_c\sin i < 80~M_J$, but
excluding objects with astrometric masses in the stellar regime (this paper).
Lower panel: cumulative [Fe/H] (with no period cut). Same symbols as in the upper panel.}
\label{f:plot3}
\end{figure}

\citet{ribas07} noted a possible trend in the eccentricity-mass distribution of giant planets, with more massive
planets ($M_c \sin i > 4~M_{J}$) orbiting on average with larger eccentricities. The recent analysis by
Wright et al. (2009), who used 1 $M_J$ as the cut-off value, confirm this trend.
The~\citet{ribas07} analysis also suggested that the metallicity of stars hosting
massive planets and brown dwarf candidates is compatible to that of the solar neighborhood ([Fe/H]$\simeq -0.15$,
e.g.~\citealt{nordstrom04}), differing at the 3-$\sigma$ level from the super-solar metallicity distribution of
stars hosting lower mass planets ([Fe/H]$\simeq+0.24$, e.g.~\citealt{fv}). Hints of such a trend
had already been seen in a more limited dataset by~\citet{rice03}.
We revisit here this statistical analysis, considering all objects included in Table~\ref{t:bd} as well
as the sample of exoplanets with $M_c\sin i < 10$ $M_J$ reported in~\citet{butler06}.
For this purpose, we have divided the full sample in three subsets according to minimum mass:
$M_c\sin i< 4$ $ M_J$, $4 < M_c\sin i< 10$ $ M_J$, and $10 < M_c\sin i< 80$ $ M_J$, plus a fourth
subsample formed by those objects in Table~\ref{t:bd} with estimated true masses in the range
$10 < M_c< 80$ $ M_J$. We have then performed on these subsets a Kolmogorov-Smirnov (K-S) test,
to measure to what extent their $e$ and [Fe/H] distributions might differ, and a Wilcoxon Rank-Sum (R-S)
test, to measure to what degree the mean $e$ and [Fe/H] might be considered different. The resulting
values of the probabilities (Pr($D$) and Pr($Z$), respectively) of the null hypothesis for both tests on the various subsamples are
reported in Table~\ref{tests}, while Figure~\ref{f:plot3} shows a comparison between the
resulting cumulative distributions of $e$ and [Fe/H] for the abovementioned subsets of systems.

\begin{table}
\centering
\caption{Results of the K-S and R-S tests on different
subsets of systems: all planets from the Butler et al. (2006) catalog with
$M_c\sin i < 4$ $M_J$
(1), companions with $4 < M_c\sin i < 10$ $M_J$ (2), companions with $10 <
M_c\sin i < 80$ $M_J$
but excluding those with confirmed mass larger than 80 $M_J$ (3),
and companions with true masses in the range $10 < M_c < 80$ $M_J$ (4) .}
\label{tests}
\begin{tabular}{@{}ccccc}
\hline
Sub-sample & $D$ & $Pr(D)$ & $Z$ & $Pr(Z)$\\
\hline
& & & & \\
\multicolumn{5}{c}{Eccentricity Distribution} \\
& & & & \\
\hline
1 vs 2     & 0.435 & 0.002 & 3.142  & 0.001 \\
1 vs 4       & 0.50  & 0.08  & 2.18   & 0.01  \\
1 vs 3+4     & 0.449 & 0.001 & 3.4535 & 0.0003 \\
2 vs 4       & 0.32  & 0.64  & 0.73   & 0.23  \\
2 vs 3+4     & 0.21  & 0.65  & 0.24   & 0.41  \\

\hline
& & & & \\
\multicolumn{5}{c}{Metallicity Distribution} \\
& & & & \\
\hline
1 vs 2      & 0.22  & 0.27  & -0.67  & 0.25  \\
1 vs 4      & 0.46  & 0.06  & -1.57  & 0.06  \\
1 vs 3+4    & 0.389 & 0.002 & -2.745 & 0.007 \\
2 vs 4      & 0.31  & 0.51  & -1.11  & 0.13   \\
2 vs 3+4    & 0.28  & 0.23  & -1.36  & 0.09  \\

\hline

\end{tabular}
\end{table}

From investigation of Table~\ref{tests} and Figure~\ref{f:plot3},
a few conclusions can be drawn. As for the $e$ distribution, the
results further corroborate the notion that eccentricities of not
so massive planets are clearly less pronounced than those of more
massive planets and brown dwarf companions\footnote{The
statistical analysis presented here does not take into account the
eccentricity bias in Keplerian fits to Doppler data collected by
radial-velocity planet search programs, which a) underestimates
the abundance of low-amplitude, low-eccentricity planets, and b)
makes low-amplitude, high-eccentricity planets hard to
unveil~\citep{shen08,otoole09,valenti09}}. On the other hand,
massive planets appear to have an overall $e$ distribution which
is indistinguishable from that of brown dwarf candidates.
Similarly, the [Fe/H] distribution of the hosts appears
significantly different for companions with $M_c\sin i < 4$ $M_J$
and for brown dwarf companions, while other trends are marginal,
and in particular metallicities of massive planet hosts and of
primaries with brown dwarf companions are statistically the
same.\footnote{It also appears that the brown dwarfs with the
largest actual mass estimates seem to be found more frequently as
close companions to stars belonging to the thick disk population
of the Milky Way (see Table~\ref{t:bd}), but this can only be
regarded as a tentative speculation, given the very limited
statistics, large mass uncertainties, and heterogeneity of the
datasets.}

\subsection{Implications for formation and structural models of massive planets and brown dwarfs}

The above evidence can be interpreted, keeping in mind the
relatively small-number statistics regime we're dealing with here,
in the context of the proposed formation scenarios and internal
structure models of high-mass planets and brown dwarfs, and as a
function of the main properties of the stellar hosts (binarity,
mass, metallicity).

\subsubsection{Formation and orbital evolution of massive planets and brown dwarfs}

As already remarked in the Introduction, the lower mass limit for the formation of self-gravitating
objects from fragmentation of molecular cloud cores is today thought to be on the order of a
few $M_J$ (e.g.,~\citealt{whit07};~\citealt{luhman07}).~\citet{ribas07} and, more recently,
~\citet{font09} proposed direct cloud fragmentation, followed by inward migration by disk capture,
as a formation mechanism for the high-mass tail of the planetary population.
The detection of a significant number of free-floating objects down to
$\sim 6$ $M_{J}$ and the observed continuity of the substellar mass function in young star forming regions
(e.g.,~\citealt{caballero07}, and references therein;~\citealt{zucksong09}, and references therein)
are indications that indeed a star-like formation process may form brown dwarfs as well as planetary mass objects
\footnote{However, see e.g.~\citet{boss00} and~\citet{bate02} for alternative scenarios in which
isolated planetary mass objects are the result of ejections through dynamical interactions in protoplanetary
disks}. Indeed, the process outlined by~\citet{font09} should be effective
independently of the metallicity of the parent cloud, and is expected
to produce an eccentricity distribution similar to that of binary systems. The results shown in Table~\ref{tests}
and in Figure~\ref{f:plot3} can be seen as supportive of this scenario.
On the other hand, the disk capture mechanism proposed by~\citet{font09} does not naturally
explain the existence of the brown dwarf desert, as in their model increasingly more massive objects
should actually be easier to capture. Also, the observed
differences in the mass distribution between isolated objects and companions to solar-type stars,
the latter, as shown by \citet{grether06}, exhibiting a minimum (the ``driest part of the brown dwarf desert'')
at $\approx 30~M_{J}$ ($31^{+25}_{-18}~M_{J}$), do not appear to be readily explainable by this mechanism.

Massive planets and brown dwarfs can also be formed by gravitational fragmentation of extended disks around
solar-type primaries. In the models of e.g.~\citet{stama09}, brown dwarfs form by direct
gravitational collapse only in the outer regions of the disks, and tend to be scattered further out, or even
into the field. This mechanism can explain the existence of the brown dwarf desert at small separations.
The two other conclusions that can be drawn from these models are that 1) no planets formed by disk instability
can really be found as close companions, as they can only form in the outer regions of the disks and are
scattered out with much higher efficiency than brown dwarfs, and b) as a direct consequence short-period
Doppler-detected massive planets and brown dwarfs should be preferentially low mass stars. On the one hand,
this mechanism can be seen as supported by the data presented here (the properties of massive planets,
brown dwarfs and low mass stellar companions being very similar). On the other hand, some of the objects
in the sample of Table~\ref{t:bd} can hardly be explained by this approach (e.g., the transiting objects
\object{XO-3}b and \object{CoRoT-3b}).

In the still theoretically debated (e.g.,~\citealt{mayer02};~\citealt{rafikov05};~\citealt{stama08};~\citealt{boss09};
for a review see~\citealt{durisen07}) disk instability model of giant planet formation, massive planets with
relatively eccentric orbits and moderate to large orbital separations are its more likely product (e.g.,~\citealt{rice03};
~\citealt{boley09}). No clear prediction of the expected shapes of the orbital elements and mass distributions of
planets formed by disk instability is yet available, due to the numerical complexities of the simulations and to
some still open theoretical issues on the input physics (see e.g.~\citealt{boss09}, and references therein).
However, if high-precision measurements of the actual masses of massive planets as well as of those of the substellar companions
listed in Table~\ref{t:bd} were to become available, and they were to prove that these objects are not
preferentially low mass stars, this evidence would argue in favor of a common formation mode for massive
planets and brown dwarfs.

In the more widely accepted core-accretion mechanism for the formation of giant planets (e.g.,
\citealt{pollack96};~\citealt{alibert05};~\citealt{ida05}; for a review see~\citealt{lissa07})
very massive planets and low-mass brown dwarfs on eccentric orbits are not a natural outcome
(e.g.,~\citealt{pollack96,alibert05,ida05,kennedy08}).
However, recent models have shown that this mechanism might also be capable of forming such massive objects.
Indeed, in the model of \cite{mord1} planets with masses as large as $38~M_{J}$ are formed in long-lived massive disks
around a solar mass star. Given the increasingly lower probability of forming very massive planets or
brown dwarfs ('deuterium-burning' planets in the words of~\citet{baraffe08}), the brown dwarf desert appears compatible with the
\citet{mord1} model (for example, the probability of forming objects with masses exceeding the deuterium-burning
threshold is found to be $\sim0.4\%$, in accord with the~\citet{marcy00} estimate).
While the \citet{mord1} model does not include multi-planet systems and planet-planet
interactions, likely to play an important role in shaping the observed mass, period and eccentricity distributions,
we note that the expected location of such very massive planets is between 2 to 5 AU, similar to that
observed for \object{HD 38529c}, \object{HD 168443c} and the object studied in this paper, \object{HD 131664b}.
However, no prediction of this model on the eccentricity distribution is available, which would be a
critical element for discriminating between this mechanism and the other aforementioned options.
In addition, super-planets and low-mass brown dwarfs do exist also at small separation, where they are not expected
on the basis of the \cite{mord1} model (e.g., the transiting systems \object{HAT-P-2b}~\citep{bakos07},
\object{WASP-14b}~\citep{joshi09}, \object{XO-3b}, and \object{CoRoT-3b}, but also other massive objects
such as \object{HD 162020b}~\citep{udry02} and \object{HD 41004Bb}~\citep{zucker04}).

Finally, of particular interest are the initial claims of a
possible correlation between massive planets, eccentric orbits,
and large values of the angle between a planetary orbit and the
stellar rotation axis, as determined from spectroscopic
measurements of the Rossiter-McLaughlin (R-M)
effect~\citep{rossiter24,mclaugh24} in transiting systems. The
\object{XO-3}, \object{WASP-14}, and \object{HD 80606} systems all
have a close-in, massive planet on a very eccentric orbit and with
a significant spin-orbit misalignment (e.g.,~\citealt{johnson09},
and references therein) \footnote{While the results are not
conclusive due to the presence of systematics in the dataset, it
is worth noting that the CoRot-3 system also exhibits a formally
non-zero spin-orbit angle~\citep{triaud09}. However the large
uncertainties reported make the detection of spin-orbit
misalignment in the CoRoT-3 system only marginal.}. At first
glance, this might imply that the orbital migration history of
massive, eccentric exoplanets is somehow different from that of
less massive close-in Jupiters, However, the picture is likely not
to be so simple, not only because of the evidence for other
transiting systems with massive, eccentric planets (HAT-P-2, HD
17156) showing no signs of spin-orbit misalignment
(e.g.,~\citealt{winn07};~\citealt{barbieri09}), but also for the
recent measurements of the R-M effect in transiting systems with
Jupiter-mass planets on circular orbits, such as CoRot-1, HAT-P-7,
and WASP-17 (~\citealt{pont09}, and references therein) which
point at large values of the projected spin-orbit angles. Overall,
the evidence collected so far is a likely indication of the
variety of possible outcomes of the complex process of migration
to close-in orbits of companions with a wide range of masses,
which include a host of proposed dynamical mechanisms (e.g.,
interactions between a planet and the gaseous/planetesimal disk,
planet-planet resonant interactions, close encounters between
planets, and secular interactions with a companion star) as well
as different formation scenarios. Statistical studies such as the
one carried out by~\citet{fabrycky09}, who showed the emergency of
a bimodal distribution of spin-orbit angles, on an increasingly
larger sample of transiting systems will ultimately be the optimal
way to compare an ensemble of measurements of the R-M effect with
the predictions of migration theories.

\subsubsection{The impact of binarity}

Among the distinctive features of exoplanets discovered around members of
various types of binary and triple systems (e.g.,~\citealt{eggen09}, and references therein),
one of the most intriguing is the evidence that solar-type stars members of multiple stellar
systems appear to be preferential hosts of the most massive planets in short-period orbits
\citep{db07}, and the fact that the planetary companions with the highest eccentricities
all have a stellar or brown dwarf companion~\citep{tamuz08}. These trends
seem to indicate that planet formation and/or migration in binaries may proceed differently
than around single stars. Indeed, theoretical studies~\citep{kley00,wu03,fabrycky07}
within the context of the core accretion model of giant planet formation
suggest that the presence of a fairly close companion significantly enhances the growth rate
and make the migration timescale of the planet shorter.
On the other hand, the massive super-planets or low-mass brown dwarfs found at separations of
about 2-3 AU orbit stars that are single or with very wide companions (\object{HD 38529} has a $0.5~M_{\odot}$
common proper motion companion at a projected separation of 12000~AU), compatible with long-lived, undisturbed
disks required to form such objects according to \citet{mord1}. Models of giant planet formation by disk instability,
however, come to opposite conclusions, with giant planet formation significantly suppressed in binaries with
separations $< 100$ AU~\citep{mayer05}. From inspection of Table~\ref{t:bd}, the abovementioned trends
seem to be supported only in part. For example, the short-period objects \object{HD 98230b} and \object{HD 283750b}
orbit one of the components of wide binaries, but this does not appear to be the case for \object{HD 162020b}.
Investigations are encouraged which would aim at verifying the possible existence of binary companions to
\object{CoRoT-3b} and \object{XO-3b}. Furthermore, the brown-dwarf candidates with the highest eccentricities do not
seem to be preferentially found in multiple systems.

\subsubsection{The role of the primary mass and metallicity}

In the core-accretion model of giant planet formation, the upper
limit on the mass of the planetary companion and the final orbital
arrangement are expected to depend on stellar mass and
metallicity. Based on arguments of protoplanetary disk size and
lifetime as a function of $M_\star$, one would expect massive
planets to be found with higher probabilities around more massive
primaries~\citep{kennedy08}, and at typically moderate to large
separations~\citep{burkert07,currie09}. Indeed, the observational
evidence indicates that, for larger stellar masses, massive
companions with $\gtrsim 10 M_{J}$ are significantly more numerous
than around solar type stars (e.g.,~\citealt{lovis07,johnson08}.
For a review see ~\citet{hatzes08}, and references therein). On
the opposite end, M dwarfs are expected to show a paucity of giant
planets~\citep{laughlin04,ida05}, which is also
observed~\citep{endl06,johnson07,bailey09}. The core accretion
mechanism also naturally predicts that super-massive planets
should not be found at all around metal-poor stars, given the lack
of material for accumulation (e.g.,~\citealt{ida05}). The
alternative disk instability mechanism is instead rather
insensitive to the values of $M_\star$ and [Fe/H] of the stellar
host (e.g.,~\citealt{boss02},~\citeyear{boss06}). As a
consequence, a qualitative prediction of this model
(e.g.,~\citealt{rice03}) is that massive planets found on moderate
to large separations around massive and/or metal-deficient hosts
would be likely to have been formed by gravitational instability.
In this respect, the data collected here and the results of the
statistical analysis, that corroborate the findings
of~\citet{ribas07}, can be read as partly supportive of the latter
view. Interestingly, the frequency of (massive) planets ($f_p$)
around intermediate-mass (mostly giant) stars appears to be rather
independent of [Fe/H] ~\citep{hatzes08}. However, the four known
planet - brown dwarf systems (\object{HD 38529}, \object{HD
168443}, \object{HD 202206}, \object{HAT-P-13}) all have
super-solar mass, very metal-rich (main-sequence) primaries. The
latter evidence might be seen as supportive of the core accretion
formation mode, while the former dataset might point to a
significant role of the disk instability mechanism. However, the
global picture is likely to rather complex. For example, Doppler
surveys for giant planets around intermediate-mass stars typically
include targets which are evolved to some degree, given that
massive main sequence stars are unsuitable for high-precision
radial-velocity measurements (too few spectral lines, often
broadened by high rotation rates). The very different dependence
of $f_p$ on [Fe/H] for intermediate-mass stars with respect to
their solar-mass counter-parts might then reflect a non-primordial
origin of the metallicity enhancement in solar-type planet
hosts~\citep{pasquini07} rather than point to different formation
modes.

As for the possible dependence of brown dwarf frequency on stellar
mass and metallicity, this is still a poorly understood issue. A
quick look at literature data allows to speculate on the
possibility that the brown dwarf desert may not be so `dry' when
it comes to close sub-stellar companions to intermediate-mass
stars, in light of a handful of systems containing at least one
companion with minimum mass in the brown dwarf regime
(e.g.,~\citealt{omiya09}, and references therein). In particular,
the recently announced system of brown dwarfs around \object{BD+20
2457} ~\citep{nied09} has an architecture very similar to that of
\object{HD 168443}, with a solar-type primary. It is reminiscent
of an origin in a massive circumstellar disk, further suggesting a
scenario in which more massive sub-stellar companions are found
around more massive stellar hosts (and their frequency also
increases with increasing primary mass). Such systems would be
more likely to have formed by local gravitational instabilities in
protoplanetary disks (e.g.,~\citealt{rice03}) rather than
protostellar cloud fragmentation, given the difficulties in
forming extreme mass-ratio binaries by the latter mechanism
(e.g.,~\citealt{bate00}). However, the combination of small-number
statistics, the different priority given to observations of stars
with very massive companions in Doppler surveys, the variable
detection thresholds as a function of companion mass and
separation due to the decreasing radial-velocity precision in
increasingly higher mass stars, and the uncertainty on the actual
mass values for most of the objects included in Table~\ref{t:bd}
(as well as those referenced above) prevents one, as of today,
from drawing any serious conclusions on the possibility that the
brown dwarf desert may move in mass and/or separation range
depending on $M_\star$ and [Fe/H].

\subsubsection{Structural and atmospheric models}

Finally, one of the most effective ways of distinguishing between
massive planets and brown dwarfs is through a comparison of their
internal structure properties and of the composition of their
atmospheres (e.g.,~\citealt{chabrier09};~\citealt{burga09}).
Studies at this level are still in their infancy status,
particularly for what concerns the possibility of determining the
actual nature of such objects based on their condensate cloud
formation properties, non-equilibrium chemistry, and atmospheric
dynamics~\citep{burga09}. Howewer, the class of transiting systems
is already providing data of relevance. For example, \citet{mord1}
have proposed that massive planets might have a large content of
heavy elements (about $0.8~M_{J}$ for the individual case shown in
their Fig.~12). In the case of the transiting massive planet
\object{HD 147506b} (HAT-P-2b), with a mass of $9.04~M_{J}$ and
$R=0.982~R_{J}$ \citep{bakos07}, just below the threshold of
$10~M_{J}$ adopted here for inclusion in Table~\ref{t:bd},
\citet{baraffe08} and \citet{leconte09} argue from the measured
radius for a total amount of heavy elements of about $1 M_{J}$,
excluding a gaseous H/He object with solar composition. This
indicates that indeed $10~M_{J}$ planets might be formed by the
core accretion mechanism, unless such a large mass is the result
of planetary collision, as speculated by \citet{baraffe08}. In the
case of the super-massive transiting planet \object{XO-3b},
despite its rather uncertain radius (due to discrepancies in the
stellar radius estimates from photometry and spectroscopy), there
seems to be no need for a large content of heavy elements in the
core \citep{winn08}). Similarly, the 22-$M_J$ transiting object
CoRoT-3b fits the \citet{baraffe08} models for solar composition,
without need for large metal enrichment~\citep{leconte09}. Both
HAT-P-2b and CoRoT-3b have surface gravities similar to those of
young, very low-mass brown dwarfs such as AB Pic (see Figure~2
of~\citealt{burga09}), suggesting an overlap in the parameter
space of some physical properties (e.g., gas pressure,
temperature). Interestingly, as discussed by~\citet{deleuil08},
all transiting massive planets orbit stars more massive than the
Sun, as predicted by the core accretion model of planet formation.
On the other hand, their actual existence on such short periods
and relatively eccentric orbits (except for CoRoT-3b), and the
fact that the metallicities of their parent stars are not skewed
towards super-solar values calls for the possibility of
differences in their origin.

In conclusion, the present-day evidence on the orbital and structural properties of massive
planets and brown dwarfs as close companions to nearby solar-type stars can be interpreted
as supportive of a picture in which different mechanisms for the formation of such objects
are at work. The above discussion also indicates how the lack of a clean, statistically significant
sample of high-mass planets and brown dwarf companions to solar-type stars with well-determined
mass estimates hampers the possibility to conclusively discriminate among the various competing
modes of formation of such objects. From an observational viewpoint, improvements in the
determination of the multiplicity properties of sub-stellar companions are very much needed, and
they will be obtained in the near future through the combined contribution of a variety of
techniques. For example, some of the massive planets and brown dwarfs objects in
Table~\ref{t:bd} are within reach of next generation direct imaging instruments
such as SPHERE \citep{sphere} and for many of them true masses will be measured with
high precision by ground-based and space-borne astrometric observatories, such as VLTI/PRIMA
and Gaia (e.g.,~\citealt{launhardt08};~\citealt{casertano08};~\citealt{sozzetti09},
and references therein). The possibility that objects with similar masses formed in different way and
exhibit different chemical composition depending on the formation mechanism open exciting perspectives
for such instruments. Therefore, they will represent key benchmarks for the calibration of the masses of substellar
objects, for furthering our understanding of the dependence of the brown dwarf desert on the properties
of the stellar hosts, and even for reaching final agreement on the actual definition of planets and
brown dwarfs themselves based not simply on semantics but rather on robust physical grounds.

\begin{acknowledgements}

This research has been partially supported by INAF through PRIN 2009
``Environmental effects in the formation and evolution of extrasolar planetary system''.
AS gratefully acknowledges support from the Italian Space Agency (Contract ASI-Gaia I/037/08/0).
We thank J.-L. Halbwachs for stimulating discussions, and the referee for a timely and useful report
which helped improve the presentation of the results. This research
has made use of the SIMBAD database, operated at CDS, Strasbourg, France,
and of NASA's Astrophysics Data System Bibliographic Services.

\end{acknowledgements}

\bibliographystyle{aa}


\end{document}